\documentclass[final,5p,times,twocolumn,authoryear]{elsarticle}

\usepackage{amssymb}
\usepackage{amsmath}
\usepackage{xcolor}
\usepackage{rotating}

\usepackage{pdflscape}
\usepackage[normalem]{ulem}

\begin{document}

\begin{frontmatter}



\title{Study of dark interactions through strong gravitational lenses} 


\author[inst1]{
F. Villalobos,
Juan Magaña,
} 
\author[inst2]{T. Verdugo}
\affiliation[inst1]{organization={Facultad de Ingeniería y Arquitectura, Universidad Central de Chile},
            addressline={Av. Francisco Aguirre 0405}, 
            city={La Serena},
            postcode={}, 
            state={},
            country={Chile}}

\affiliation[inst2]{{Instituto de Astronomía, Observatorio Astronómico Nacional, Universidad Nacional Autónoma de México},
            addressline={}, 
            city={Ensenada, B.C.},
            postcode={106}, 
            state={},
            country={México}}
\begin{abstract}

The possible interaction between the dark components of the Universe (dark matter and dark energy) stands as an attractive alternative to the standard $\Lambda$CDM cosmological model. In this work, we present a novel analysis of three sign-changeable interaction models whose coupling term $Q$ depends explicitly on the deceleration parameter $q$ and is proportional to different energy densities: dark matter, dark energy, and total energy density. To constrain these models, we combine strong gravitational lensing data on two complementary scales: a sample of early-type galaxies acting as lenses and the galaxy cluster Abell 1689.

Our results show that the interaction strength $\beta$ depends on the choice of the coupling term $Q$, with all models yielding negative values of $\beta$, indicative of a dark interaction scenario. The $\beta$ values obtained in this work are significantly larger in magnitude than those previously reported using Type Ia supernovae, CMB, and BAO. The strong-lensing constraints indicate a transition to cosmic acceleration at earlier redshifts ($z_t \sim 1.8-2.1$) than that predicted by the $\Lambda$CDM model, while remaining consistent with cosmic chronometer measurements within the reconstructed confidence regions. Therefore, our study shows that strong gravitational lensing data provide an independent and competitive cosmological probe capable of testing interacting dark energy scenarios. The sensitivity of lensing observables to the expansion history enables access to complementary information about dark-sector dynamics beyond standard cosmological probes.

\end{abstract}




\end{frontmatter}



\section{Introduction}
\label{introduction}

In the current standard cosmological model within general relativity, it is well known that there is evidence for the accelerated expansion of the Universe sourced by a repulsive dark energy (DE) component as a cosmological constant which is related to the quantum vacuum energy \citep{riess_SN1a_1998, perlmutter_SN1a_1999}. On the other hand, the model also postulates the existence of a cold dark matter (DM) component as responsible for the large-scale structure formation. 

In this model, known as the $\Lambda$ cold dark matter ($\Lambda$CDM) model, dark matter accounts for approximately $27\%$ of the total energy content of the Universe, while dark energy contributes about $68\%$; the remainder corresponds to baryonic matter. Although this paradigm is strongly supported by  several cosmological observations such as the cosmic microwave background radiation \citep[CMB, ][]{Planck_2018} and Type Ia supernovae \citep[SNeIa][]{Scolnic:2021amr} cosmic chronometers \citep{moresco_CC_2016}, among others, it still presents several open issues. For instance, the so-called \emph{coincidence problem} \citep{Copeland2006, Bamba2012} which arises from the near equality of the present-day energy densities of DM and DE, despite their different physical nature and evolutionary histories. Recently, another challenge faced by the $\Lambda$CDM model is the Hubble tension, which refers to the discrepancy between the estimations of the so-called Hubble constant ($H_0$) inferred from late and early universe observations. There are rigorous methods used to measure $H_0$, which represents the expansion rate of the Universe. Among the most precise measurements are those using SNeIa, which act as standard candles by relating their intrinsic luminosity to the observed distance \citep{riess_SN1a_1998, perlmutter_SN1a_1999}. Additionally, $H_0$ can be inferred from early-Universe data particularly from anisotropies in the CMB\citep{Planck_2018}.
While \citet{Planck_2018} report a value of $H_0=67.37\pm0.54$km/s/Mpc, the late data reports $H_0=73.30\pm1.04$km/s/Mpc \citep{Riess:2022ApJ}, indicating a tension of $\sim 5\sigma$ which is not satisfactorily explained by the $\Lambda$CDM model. 

To address these issues, a wide variety of alternative models have been proposed, including dynamical DE scenarios based on specific parametrizations, as well as modifications of gravity  \citep[see][for a review]{Motta:2021hvl, CosmoVerseNetwork:2025alb}, quantum-gravity inspired alternatives to a strictly constant $\Lambda$ \citep[e.g.,][]{2023JCAP...10..047D}, and holographic dark-energy scenarios based on generalized horizon entropies  \citep[e.g.,][]{2026NuPhB102517368S,2026arXiv260315178H}. These approaches provide complementary routes to late-time acceleration and have recently been confronted with late-time observations. Recent DESI results have also renewed interest in Quintom dynamical dark-energy models, where the effective equation of state may cross the cosmological-constant boundary \citep{2025arXiv250524732C}. However, the interpretation of these trends remains under discussion, particularly in connection with local $H_0$ determinations and possible dataset-dependent effects in BAO measurements \citep{2022JCAP...04..004L, PhysRevD.110.123519, 2026PDU....5202268O}. Another interesting possibility is that DM and DE interact with each other, i.e, there is an energy transfer from one dark component to the other through an interacting term $Q$.
These interacting dark energy (IDE)  scenarios \citep[see][for a review]{Bolotin:2013jpa} offer a plausible mechanism to explain why the densities of both components are comparable today, making the study of such models relevant and could relax the Hubble tension \citep{diValentino:2017}. 
An interesting kind of IDE models is the one where there is a change of direction in the energy transfer, i.e. a change in sign in the dark interaction, during the cosmic evolution. In an early work, \citet{wei_interaction_2011} proposed models with a sign-changeable interaction parameterized by the deceleration parameter and proportional to the energy densities of the cosmic dark fluids, and tested them using several cosmological data. Later, these models have been also studied from the dynamical system approach and using statefinder diagnostic tool parameters \citep{Arevalo:2019axj, Carrasco:2023imi}.

On the other hand, strong gravitational lensing at different astrophysical scales constitutes a competitive tool for constraining alternative cosmological models. At galaxy-cluster scales, the reconstruction of cluster mass distributions has been employed to simultaneously constrain cosmological parameters across a variety of models \citep{jullo_SL_2010,2015ApJ...813...69M, Magana2018ApJ, Caminha:2021iwo}, whereas at galactic scales, strong lensing observations have been used to constrain dynamical dark energy scenarios with different parameterizations \citep{amante_SL_2020}. Recently, \citet{verdugo_EPJC_2024} developed a new methodology to constrain the parameters of a braneworld dark energy model by simultaneously reconstructing the galaxy cluster Abell 1689 and fitting a theoretical distance ratio of a sample of strong lensing galaxies, finding that this approach provides a competitive and robust tool for testing alternative cosmological scenarios.

In the present work, following the approach proposed \citet{verdugo_EPJC_2024}, we aim to test three models of the interacting dark sector, characterized by a sign-changeable coupling function $Q$. To this end, we combine two complementary strong-lensing constraints operating at different physical scales: the mass distribution of the well-known galaxy cluster Abell~1689 \citep{jullo_SL_2010}, and the lensing properties of early-type galaxies acting as strong lenses \citep{amante_SL_2020}. 
The outline of the manuscript is as follows: Sec. \ref{sec:Model Q} is dedicated to presenting the mathematical formalism of on sign-changeable IDE models Sec. \ref{sec:methodology} describes the data and methodology used to constrain the IDE model. In Sec. \ref{sec:Results}, we show our results and finally in Sec. \ref{sec:Conclusions} we give our conclusions.

\section{Cosmological Framework}

In the following, we present the cosmological models to be tested, considering three different interaction terms. We assume a flat universe governed by the Friedmann-Robertson-Walker (FRW) metric.

\subsection{The interacting Q model}
\label{sec:Model Q}

In this work, we present parameter estimations for interacting models with coupling term $Q$ proposed by \cite{wei_interaction_2011}, in which the interaction is proportional either to the energy density of a dark fluid or to the total energy density, and depends on the deceleration parameter $q$. This approach is motivated by evidence of a change in $Q$ within the range $0.45 \lesssim z \lesssim 0.9$ \citep{Cai_Su_2010}, where the models were tested against SNe, BAO, CMB, and Hubble parameter data by dividing the redshift range into several bins and assuming a constant interaction term in each bin, leading to evidence for a sign change of $Q$ across those intervals. The model equations follow the standard interacting dark-energy framework:
\begin{eqnarray}
\dot{\rho}_{dm}+3H\rho_{dm} &=& Q, \nonumber\\
\dot{\rho}_{de}+3H(\rho_{de}+p_{de}) &=& -Q,
\end{eqnarray}

where $\rho_{dm}$  and $\rho_{de}$ are the DM and DE  energy densities.  
With this premise, the sign-changeable interaction terms are defined as
\begin{align}
Q &= q\left( \alpha \dot{\rho}_m + 3\beta H \rho_{dm} \right), \\
Q &= q\left( \alpha \dot{\rho}_{\text{tot}} + 3\beta H \rho_{\text{tot}} \right), \\
Q &= q\left( \alpha \dot{\rho}_{\text{DE}} + 3\beta H \rho_{\text{DE}} \right).
\end{align} 
where $\alpha$ and $\beta$ are dimensionless constants. 
We consider only the case $\alpha = 0$, which preserves the key physical feature of sign-changing interactions while admitting analytical solutions. The general case $\alpha \neq 0$ leads to intrinsically nonlinear differential equations that lack closed-form solutions and require computationally intensive numerical methods beyond the scope of this analysis. Notice that $Q$ can change the direction of the energy flow when $q$ change of sign, i.e. when the Universe passes from a decelerated stage to an accelerated one.

\subsection{Case $ Q = q (3\beta H \rho_{dm} ) $ }
For the parameterization $ Q = 3\beta q H \rho_{dm} $, the dimensionless Friedmann equation becomes

\begin{equation}
\begin{aligned}
E(z) \equiv \frac{H}{H_0} 
&= \left\{ 1 - \frac{2 + 3\beta}{2(1 + \beta)} \, \Omega_{dm0} \right. \\
&\quad \left. \times
\left[ 1 - (1 + z)^{3(1 + \beta)} \right] 
\right\}^{\frac{1}{2 + 3\beta}},
\end{aligned}
\label{Ecaso1}
\end{equation}
being $\Omega_{dm0}=\Omega_{dm}(z=0)$. The deceleration parameter defined as $q \equiv -\frac{\ddot a}{aH^2}$ can be written in terms of redshift and $E(z)$ as
\begin{equation}
q(z)= \frac{(1+z)}{E(z)}\frac{dE(z)}{dz}-1.
\label{eq:qz}
\end{equation}
Thus, the analytical expression for $q$ reads as 
\begin{equation}
q(z) = -1 + \frac{3\Omega_{dm0}(1+z)^{3(1+\beta)}}{2\left[1 - \dfrac{2+3\beta}{2(1+\beta)}\Omega_{dm0}\left(1 - (1+z)^{3(1+\beta)}\right)\right]} .
\label{eq:qi}
\end{equation}

Which allows us to track the evolution of the cosmic expansion history.

\subsection{Case $Q = q(  3\beta H \rho_{\text{tot}} ) $ }

For the interaction $Q = 3\beta q H \rho_{\text{tot}}$, the the dimensionless Friedmann equation is determined by

\begin{equation}
E(z) = (1 + z)^{3(2 - 3\beta + r_1)/8} \,
\left[ \frac{(1 + z)^{-3r_1/2} + C_{21}}{1 + C_{21}} \right]^{1/2} ,
\label{Ecaso2}
\end{equation}

where $r_1 = \sqrt{4 + \beta (4 + 9\beta)}$ and the integration constant $C_{21}$ is given by:

\begin{equation}
C_{21} = -1 + \frac{2 r_1}{2 - 3\beta - 4 \Omega_{dm0} + r_1} .
\label{eq:caso2_C21}
\end{equation}

For the model with interaction proportional to the total energy density, the deceleration parameter $q(z)$ is given by:

\begin{equation}
\begin{aligned}
q(z) = & -1 + \frac{3(2 - 3\beta + r_1)}{8} \\
& - \frac{3r_1}{4} 
\frac{(1+z)^{-3r_1/2}}{(1+z)^{-3r_1/2} + C_{21}} .
\end{aligned}
\end{equation}

Using the definitions of $r_1$ and $C_{21}$ provided above.

\subsection{Case $ Q =  q(3\beta H \rho_{\text{DE}} ) $ }

For the interaction $Q = 3\beta q H \rho_{DE}$, where the coupling depends on the dark energy density, the dimensionless Friedmann equation is determined by

\begin{equation}
\begin{aligned}
E(z) \equiv \frac{H(z)}{H_0} 
&= (1 + z)^{\frac{3(2 - 5\beta + r_2)}{4(2 - 3\beta)}} \\
&\quad \times
\left[
\frac{(1 + z)^{-3r_2/2} + C_{31}}{1 + C_{31}}
\right]^{\frac{1}{2 - 3\beta}} .
\end{aligned}
\label{eq:Ecaso3}
\end{equation}

where the parameter $r_2$ is defined as
\begin{equation}
r_2 \equiv \sqrt{(2 - \beta)^2} = |2 - \beta|,
\label{eq:caso3_r2}
\end{equation}

and the integration constant $C_{31}$, determined by the condition $\Omega_{dm}(z=0) = \Omega_{dm0}$, is

\begin{equation}
C_{31} = -1 + \frac{2r_2}{2 - 5\beta + r_2 + 2\Omega_{dm0}(3\beta - 2)}.
\label{eq:caso3_C31}
\end{equation}

For the model with interaction proportional to the dark energy density, the deceleration parameter $q(z)$ is given by:

\begin{equation}
\begin{aligned}
q(z) = & -1 + \frac{3(2 - 5\beta + r_2)}{4(2 - 3\beta)} \\
& - \frac{3r_2}{2(2 - 3\beta)}
\frac{(1+z)^{-3r_2/2}}{(1+z)^{-3r_2/2} + C_{31}} ,
\end{aligned}
\end{equation}

where $r_2$ and $C_{31}$ are defined above.

\subsection{The Equation of State}
\label{sec:weff}

We introduce the effective equation of state of the total cosmic fluid, defined as
\begin{equation}
w_{\rm eff}(z) \equiv \frac{p_{\rm tot}}{\rho_{\rm tot}} = \frac{2q(z)-1}{3},
\label{eq:weff}
\end{equation}
where $p_{\rm tot}$ and $\rho_{\rm tot}$ are the total pressure and energy density of the Universe, and $q(z)$ denotes the deceleration parameter, which depends only on the Hubble expansion history $H(z)$ \citep{2008RPPh...71e6901L}. Recent kinematic reconstructions based on $q(z)$ have used $w_{\rm eff}(z)$ and related diagnostics to test departures from the standard $\Lambda$CDM expansion history in a model-independent way \citep{2026arXiv260423992V}.

In non-interacting cosmological models, such as $w$CDM, $w_{\rm eff}(z)$ reduces to a simple weighted contribution of the dark energy component, namely $w_{\rm eff}(z)=w_X\,\Omega_X(z)$ (neglecting radiation), being $w_X$ the dark energy EoS. However, in interacting dark sector scenarios, the presence of an energy transfer term $Q$ modifies the evolution of both dark matter and dark energy. Consequently, $w_{\rm eff}$ no longer represents a single component, but rather the integrated dynamic response of the expansion rate to the interaction.

Despite these different physical origins, $w_{\rm eff}$ serves as a robust, model-independent bridge for comparison. Since gravitational lensing observables depend on integrals of $H(z)$, this parameter allows us to project the effects of an interaction onto a kinematic quantity that is directly comparable to the expansion history of a non-interacting $w$CDM model.

Thus, as a reference scenario, in this work we also consider a spatially flat, non-interacting $w$CDM cosmology, characterized by a constant dark energy equation of state $w_X$. In this case, the dimensionless Hubble parameter is given by
\begin{equation}
E^2(z) = \Omega_{dm0}(1+z)^3 + (1-\Omega_{dm0})(1+z)^{3(1+w_X)},
\end{equation}
where radiation is neglected due to the low-redshift  of the strong lensing sample. By applying Eq.~(\ref{eq:weff}) to both cases, we can consistently compare how the interacting and non-interacting scenarios drive the cosmic expansion using the same strong lensing data.

\section{Data and methodology}\label{sec:methodology}

In this work, we use gravitational lensing data of early-type galaxies and one galaxy cluster, Abell 1689.

\subsection{Strong Lensing in galaxies}

We employ the compilation of early-type galaxies acting as strong gravitational lenses presented by \citet{amante_SL_2020}. The fiducial sample comprises \( N_{\mathrm{SL}} = 143 \) strong lensing systems (SLS), each characterized by four measured quantities: the stellar velocity dispersion (\( \sigma \)), the Einstein radius (\( \theta_E \)), the lens redshift (\( z_l \)), and the source redshift (\( z_s \)).

We can constrain cosmological parameters minimizing the chi-square function given as
\begin{equation}
\chi_{\mbox{Gal}}^2 = \sum_{i=1}^{N_{SL}} \frac{ \left[ D^{th}\left(z_{l}, z_{s}; \bf{\Theta_{Cos}} \right)  -D^{obs}(\theta_{E},\sigma^2)\right]^2 }{ (\delta D^{\rm{obs}})^2},
\label{eq:chisquareSL}
\end{equation}
where we define the ratio of two angular diameter distances $D \equiv D_{ls}/D_{s}$. The theoretical ratio $D^{th}$ is calculated from Eqs. (\ref{Ecaso1}), (\ref{Ecaso2}), and (\ref{eq:Ecaso3}), using the definition of the angular diameter distance. On the other hand, through the observationally measured properties, we get the observed counterpart $D^{obs}$ , and $\delta D^{\rm{obs}}$  as the error propagation of the $D^{obs}$ function.  The vector $\mathbf{\Theta_{Cos}}$ is formed by the parameters $\beta$ and $\Omega_{dm0}$.

\subsection{Strong lensing in Abell 1689}

Following \cite{jullo_SL_2010}, to reconstruct the A1689 mass model and simultaneously  constrain the cosmological parameters  we use the 'family ratio' which is defined  as the angular diameter distance ratios of two images from different sources. Our model uses 28 images from $N_f$ = 12 families, all with measured spectroscopic redshifts in the range $1.15< z_{\rm s} < 4.86$. For the present work, we adopt an error $\Delta^2$ = 0.5" in the position of the images. In this case, the $\chi^{2}$ for a multiple image system $i$ is defined as 
\begin{equation}\label{eq:Chi2Lens}
\chi_{i}^{2} = \sum_{j=1}^{n_i}
\frac{\left| \vec{x}_{\rm obs}^j - \vec{x}^j(\mathbf{\Theta_{W}}) \right|^2}{\Delta^{2}}\;,
\end{equation}
where $n_i$ is the number of multiple images for the source $i$, $\vec{x}_{\rm obs}^j$ is the observed position corresponding to image $j$, and $\vec{x}^j(\mathbf{\Theta_{W}})$ is the position of image $j$ predicted by the  model, whose total parameters (the cosmological parameters and the cluster parameters) are included in the
vector $\mathbf{\Theta_{W}}$. Thus, $\chi^2_{\mbox{Clu}} =\sum_{1}^{N_f} \chi_{i}^{2}$, is the total chi square function in this case.

\subsection{Cosmological parameter estimation}

Assuming a Gaussian likelihood $\mathcal{L} \propto e^{-\chi^{2}_{\mathrm{Tot}}/2}$, the total chi-square function is defined as
\begin{equation}\label{eq:Chi2Tot}
\chi_{\mathrm{Tot}}^{2} = \chi^2_{\mathrm{Gal}} + \chi^2_{\mathrm{Clu}},
\end{equation}
where $\chi^{2}_{\mathrm{Tot}}$ represents the combined contribution of both complementary strong-lensing approaches: the sample of early-type galaxies acting as lenses and the cluster-scale constraints from Abell~1689. This total chi-square is minimized within {\sc LENSTOOL} \citep{Jullo2007} using a Bayesian Markov Chain Monte Carlo (MCMC) algorithm, following the methodology described in \citet{verdugo_EPJC_2024}, in order to obtain the posterior median values of the cosmological model parameters. Our Bayesian analysis employs a $3\sigma$ Gaussian prior on the dark matter density parameter $\Omega_{dm0} = 0.265 \pm 0.005$ following \citet{Planck_2018}, and a uniform prior $\beta \in [-1, 1]$ for the interacting parameter. During the sampling, we discard any parameter set that leads to a non-real
background evolution, namely whenever the reconstructed dimensionless
expansion rate squared satisfies $E^2(z)<0$ for any redshift entering the
likelihood.(see~\ref{App1}).

\subsection{Model selection}

To compare the different interaction models, we will use the Akaike information Criterion (AIC; \citealp{akaike1974}) and the Bayesian information Criterion (BIC; \citealp{schwarz1978}), defined as:
\begin{equation}
\mathrm{AIC} = \chi^{2}_{med} + 2k ,
\end{equation}

\begin{equation}
\mathrm{BIC} = \chi^{2}_{med} + k \ln N ,
\end{equation}
where $\chi^{2}_{med}$ is the chi-square value obtained from the posterior median of the parameters, $k$ is the number of parameters, and $N$ corresponds to the number of data points used in the fit. In this case,
$k=21$, accounting for the cluster mass–profile parameters and the free parameters of the model, and $N=171$, corresponding to the total sample used in the analysis.

Since AIC and BIC are meaningful only in a relative sense, the comparison is made by calculating the differences with respect to the model with the lowest (best) value. For a given model $i$, we define:

\begin{equation}
\Delta \mathrm{AIC}_i = \mathrm{AIC}_i - \mathrm{AIC}_{\min}
\end{equation}

\begin{equation}
\Delta \mathrm{BIC}_i = \mathrm{BIC}_i - \mathrm{BIC}_{\min}
\end{equation}

Following standard conventions, models with $\Delta\mathrm{AIC}\leq2$ have substantial empirical support, values in the range $4$--$7$ indicate considerably less support, and $\Delta\mathrm{AIC}>10$ implies essentially no support. For the BIC, $\Delta\mathrm{BIC}\leq2$ indicates negligible evidence against the model, $2$--$6$ positive evidence, $6$--$10$ strong evidence, and values larger than $10$ very strong evidence against the model.

\section{Results and Discussion}\label{sec:Results}
In the following, we present our results for the three cosmological models.

\begin{figure}[htbp]
    \centering
    \begin{minipage}{0.4\textwidth}
        \centering
        \includegraphics[width=\textwidth]{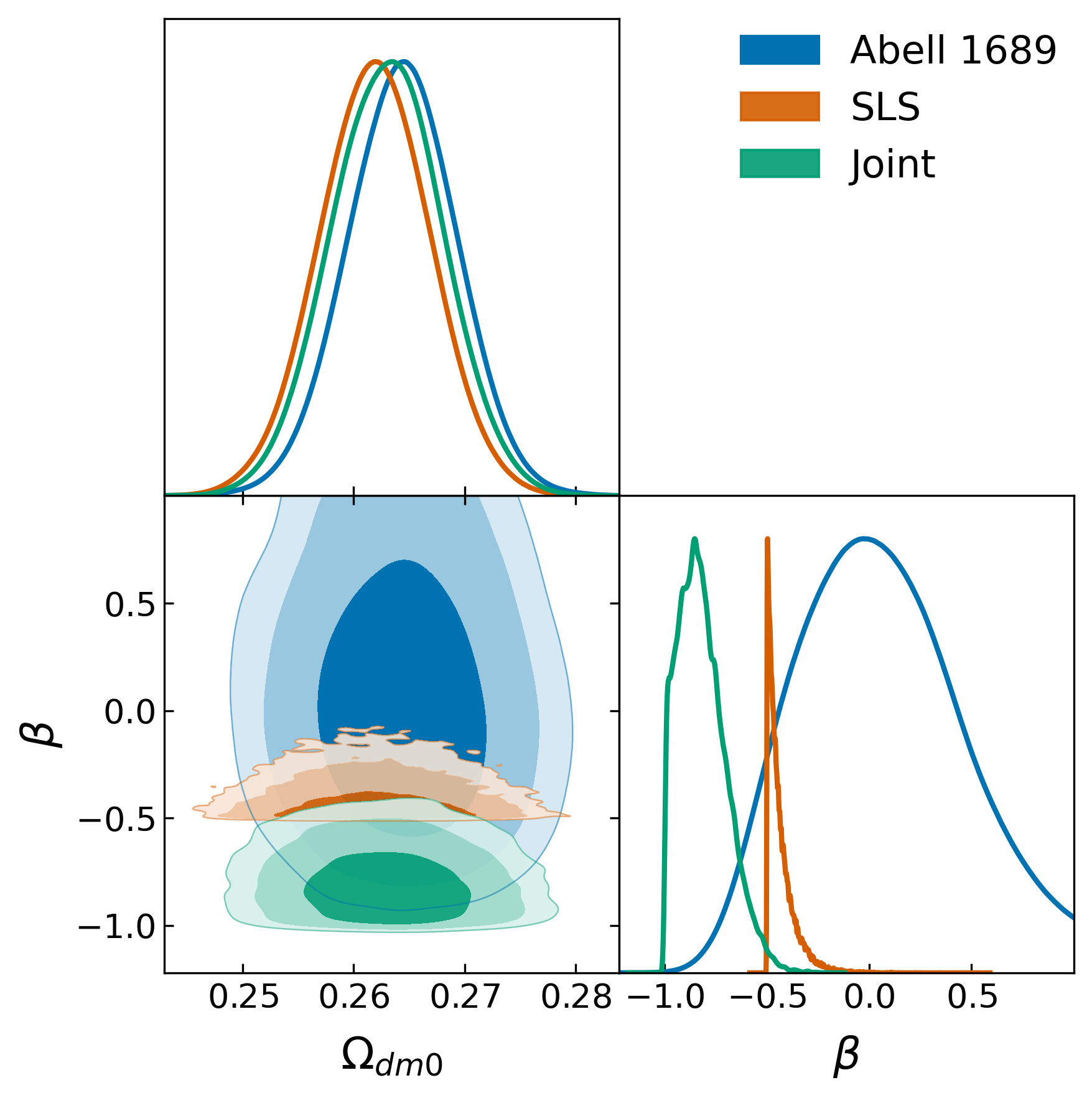}
    \end{minipage}\hfill
    \begin{minipage}{0.4\textwidth}
        \centering
        \includegraphics[width=\textwidth]{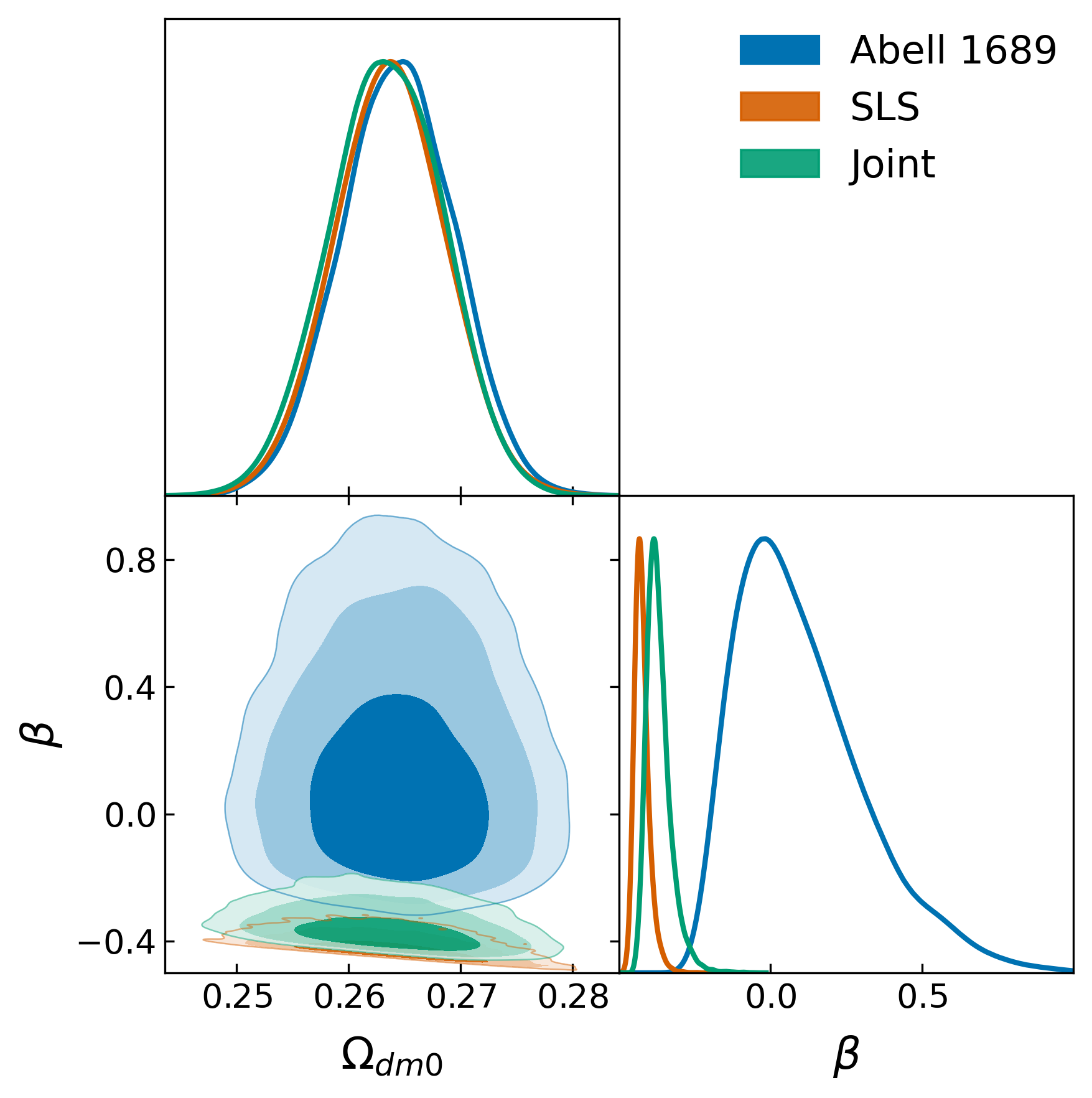}
    \end{minipage}\hfill
    \begin{minipage}{0.4\textwidth}
        \centering
        \includegraphics[width=\textwidth]{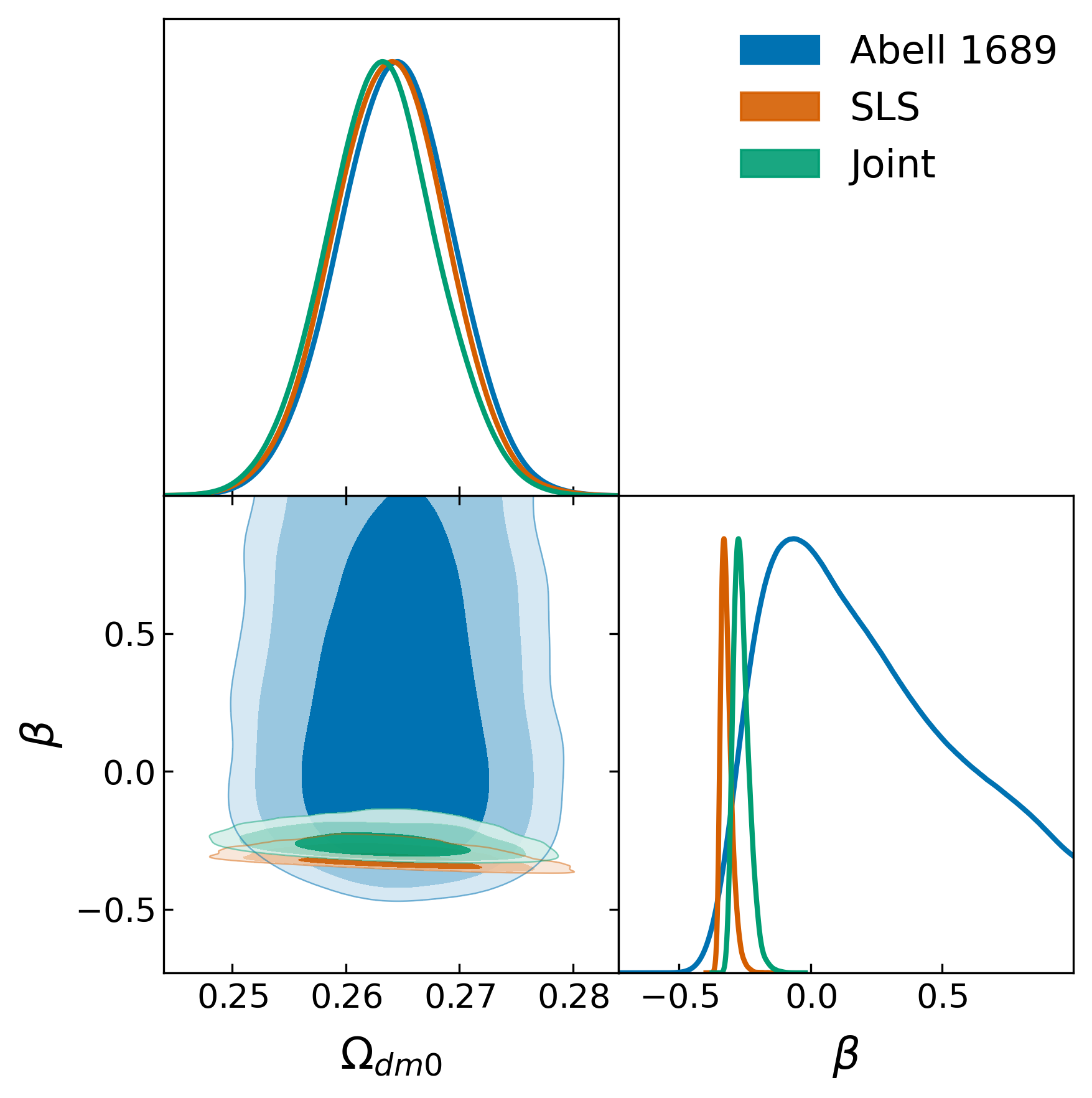}
    \end{minipage}

    \caption[MCMC distributions for interacting dark energy models]{
    Joint posterior distributions for the parameters \( \Omega_{dm0} \) and \( \beta \) for three interacting dark energy models.
    The top, middle, and bottom panels correspond to interactions proportional to the matter density, total energy density, and dark energy density, respectively. The contours represent the 68\%, 95\%, and 99\%  regions.
    }
    \label{fig:triangle_all}
\end{figure}

\begin{figure*}\begin{center}
\includegraphics[scale=0.4]{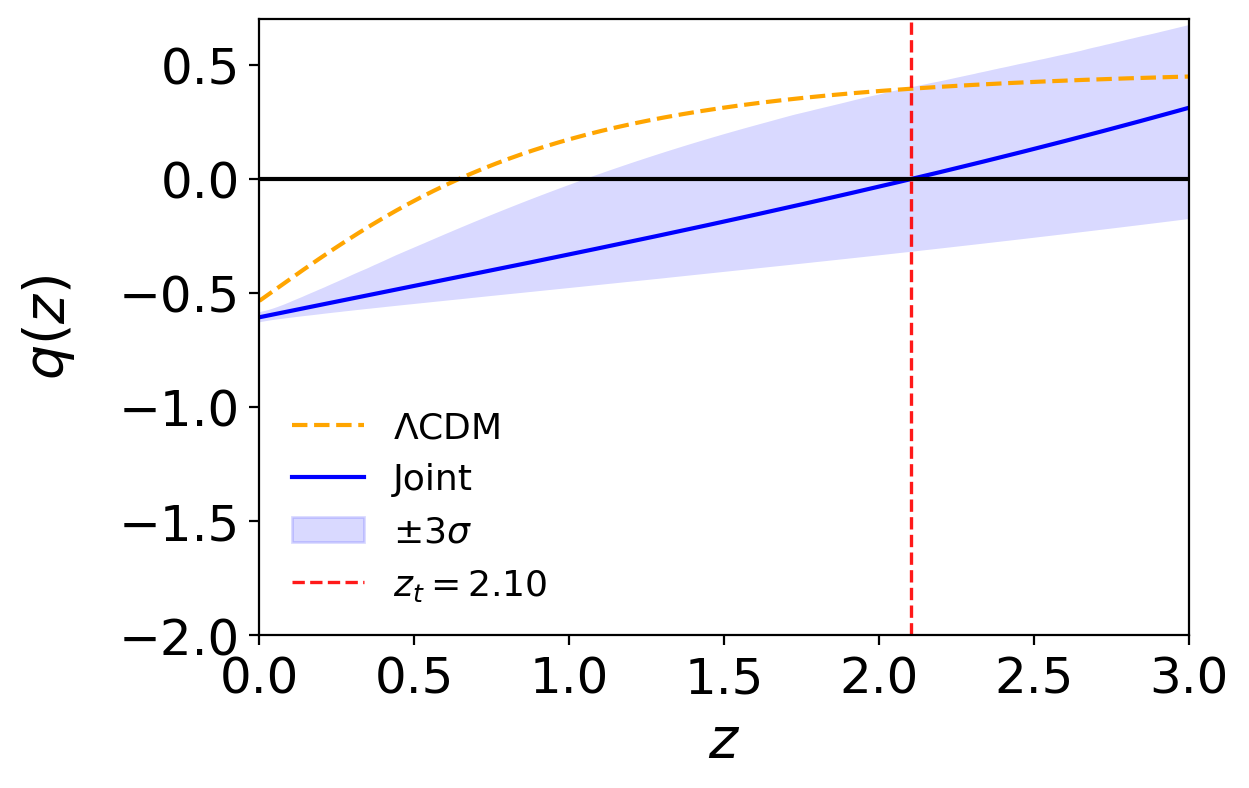}
\includegraphics[scale=0.4]{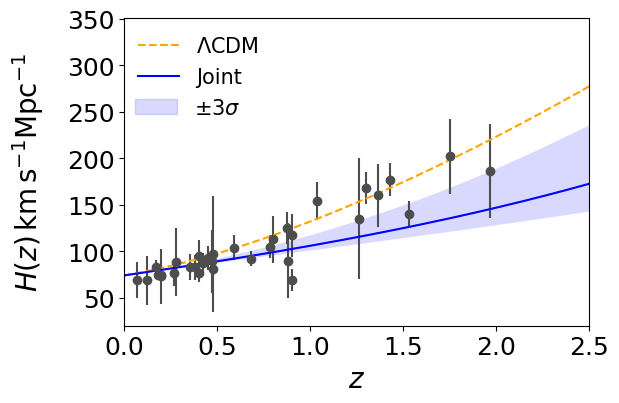}\\
\includegraphics[scale=0.4]{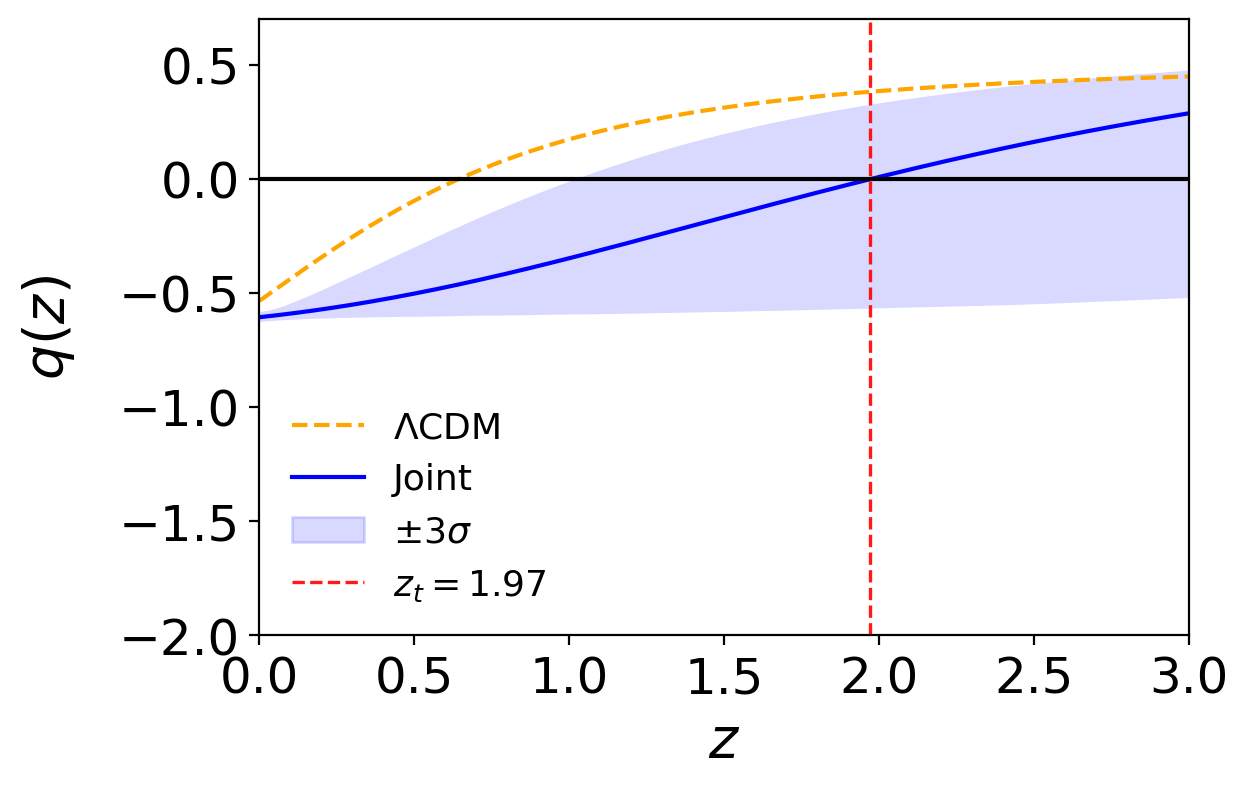}
\includegraphics[scale=0.4]{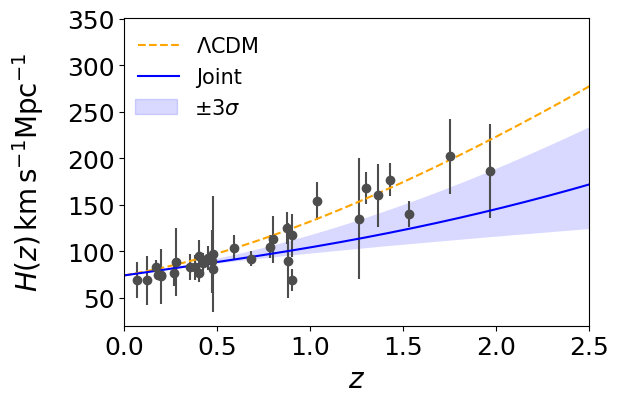}\\
\includegraphics[scale=0.4]{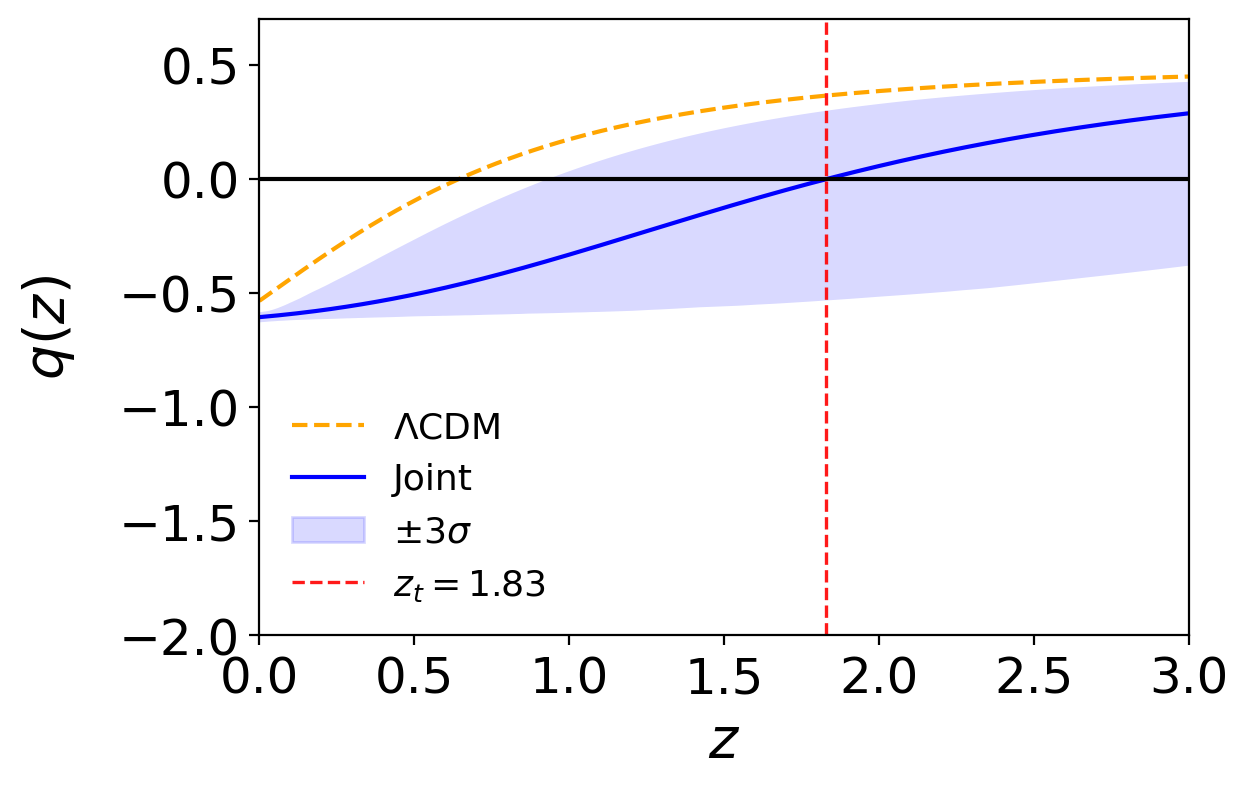}
\includegraphics[scale=0.4]{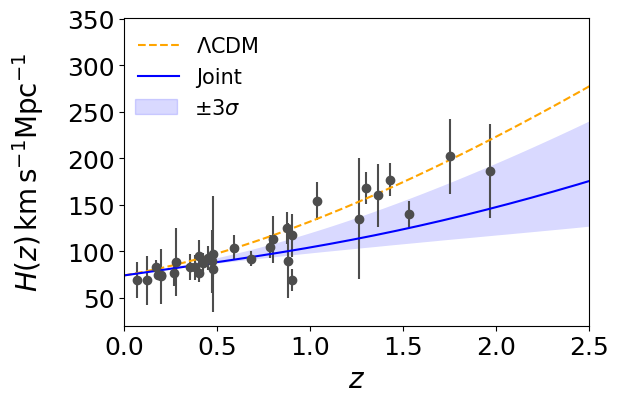}\\
\caption{Evolution of the deceleration parameter $q(z)$ (left panels) and the Hubble parameter $H(z)$ (right panels)
for three interacting dark energy models.
Panels top, middle, and bottom correspond respectively to
$Q = q(3\beta H \rho_{dm})$,
$Q = q(3\beta H \rho_{\text{tot}})$, and
$Q = q(3\beta H \rho_{\text{DE}})$.
Solid lines denote the interacting scenarios, dashed lines the $\Lambda$CDM model. The vertical red dashed line indicates the transition redshift $z_t$.
Shaded regions represent the $3\sigma$ confidence intervals.
Cosmic chronometer data are taken from \cite{moresco_CC_2016}.}

\label{fig:combined_qz_Hz}
\end{center}\end{figure*}

\subsection{Model $Q = q (3 \beta H \rho_{dm})$}

For this model, the joint SL
constraints lead to a cosmological scenario that departs significantly from
the standard $\Lambda$CDM expansion history.

The joint strong-lensing MCMC analysis yields the posterior distributions displayed in Fig.~\ref{fig:triangle_all} (top panel), with best-fit values $\Omega_{dm0} = 0.2631^{+0.0050}_{-0.0051}$ and $\beta = -0.8290^{+0.1334}_{-0.1053}$. The interaction parameter is strongly negative and significantly different
from zero, indicating an intense energy transfer within the dark sector.

Compared to the results reported by \citet{wei_interaction_2011}, who found
$\Omega_{dm0} \simeq 0.27$ and a weakly negative interaction
($\beta \sim -0.01$) using SNe~Ia, CMB, and BAO data, our lensing-based
constraints favor a substantially stronger interaction.
This suggests a more efficient transfer of energy from dark energy to dark
matter than previously inferred from other background observables.

The inferred matter density parameter is also marginally lower than $\Omega_{dm0} = 0.265 \pm 0.005$ by \citet{Planck_2018}, while higher
than the value obtained by \citet{wei_interaction_2011}.

The reconstructed $q(z)$ parameter, shown in the left column of
Fig.~\ref{fig:combined_qz_Hz} (top panel), illustrates the dynamical impact of the
strongly negative interaction parameter.
The transition to cosmic acceleration, defined by $q(z_t)=0$, occurs at
$z_t = 2.10$, significantly earlier than in the $\Lambda$CDM model, for which
$z_t \simeq 0.64$ \citep{Planck_2018}.

The corresponding expansion history $H(z)$, displayed in the right panel of
Fig.~\ref{fig:combined_qz_Hz} (top panel), predicts a systematically lower expansion rate
than $\Lambda$CDM at intermediate and high redshifts, while remaining
consistent with cosmic chronometer measurements within the $3\sigma$
confidence region.
This behavior reflects the effect of the interaction term could reduce the Hubble tension
\citep{DiValentino2021,Riess:2022ApJ}.


\subsection{Model $Q = q(  3\beta H \rho_{\text{tot}} ) $}

The middle panel of Fig. \ref{fig:triangle_all} shows the strong-lensing posterior distributions for this model.
The best-fitting parameters  are
$\Omega_{dm0} = 0.2632^{+0.0049}_{-0.0049}$ and
$\beta = -0.2662^{+0.0343}_{-0.0253}$.

In comparison with the $Q \propto \rho_{dm}$ model, the DM density parameter is slightly higher, although it remains below the Planck estimate. The best-fit for $\beta$ is significantly smaller in magnitude than in the $Q \propto \rho_{dm}$ scenario. Nevertheless, $\beta$ remains significantly stronger than that ($\beta \sim -0.01$) reported by \citet{wei_interaction_2011}.

The reconstructed $q(z)$ parameter, shown in the middle-left panel of Fig. \ref{fig:combined_qz_Hz}, reveals that
the transition to the accelerated phase still occurs at a higher redshift, $z_t = 1.97$, compared to $z_t \simeq 0.64$ of the standard model.
This behavior indicates that coupling the interaction to the total energy
density $\rho_{\mathrm{tot}}$ can trigger the onset of cosmic
acceleration, even for moderate values of $\beta$.

The corresponding expansion history $H(z)$, displayed in the middle-right panel of
Fig.~\ref{fig:combined_qz_Hz}, is similar to the
$Q \propto \rho_{dm}$ model.
In both cases, the interacting scenario predicts a systematically lower
expansion rate than $\Lambda$CDM at intermediate redshifts
($1 \lesssim z \lesssim 3$), while remaining consistent with cosmic
chronometer measurements within the $3\sigma$ confidence level.

\subsection{Model $Q = q(  3\beta H \rho_{\text{DE}}) $}

For this model, we obtain the joint strong-lensing constraints
$\Omega_{dm0} = 0.2634^{+0.0051}_{-0.0052}$ and
$\beta = -0.3756^{+0.0430}_{-0.0302}$
(see the bottom panel of  Fig.\ref{fig:triangle_all}). This $\Omega_{dm0}$ value is consistent with the other cases but slightly below the Planck 2018 estimate  \citep{Planck_2018}.

The $\beta$ parameter estimationis significantly more negative than $\beta \sim -0.01$ obtained by \citet{wei_interaction_2011} using SNe~Ia, CMB, and BAO.
Its magnitude lies between that obtained for the models coupled to
$\rho_{dm}$ ($\beta = -0.8928$) and to $\rho_{\rm tot}$ ($\beta = -0.3201$),
indicating an intermediate interaction strength.

Fig.~\ref{fig:combined_qz_Hz} (bottom-left panel) illustrates the reconstructed deceleration parameter $q(z)$ pointing put a transition redshift $z_t = 1.83$.
Although this value is lower than those obtained for the models coupled
to $\rho_{dm}$ ($z_t = 2.10$) and $\rho_{\rm tot}$ ($z_t = 1.97$),
it remains substantially higher than the
$\Lambda$CDM prediction ($z_t = 0.64$).
This decrease of $z_t$ from $\rho_{dm}$ to $\rho_{\rm tot}$ and
finally to $\rho_{\rm DE}$ suggests a  change in the efficiency with which the different couplings modify the expansion dynamics \citep{2019MNRAS.482.1858Y, 2025PDU....4801951G}.

The corresponding expansion history $H(z)$, displayed in the bottom-right panel of Fig.~\ref{fig:combined_qz_Hz}
is similar to that obtained for the other interacting
models.
At intermediate redshifts ($1 \lesssim z \lesssim 3$), there is a slight deviation to the $\Lambda$CDM model prediction. Nevertheless, the results remain consistent with cosmic chronometer measurements within $3\sigma$ confidence level.

\begin{figure}[htbp]
\centering
\includegraphics[width=0.45\textwidth]{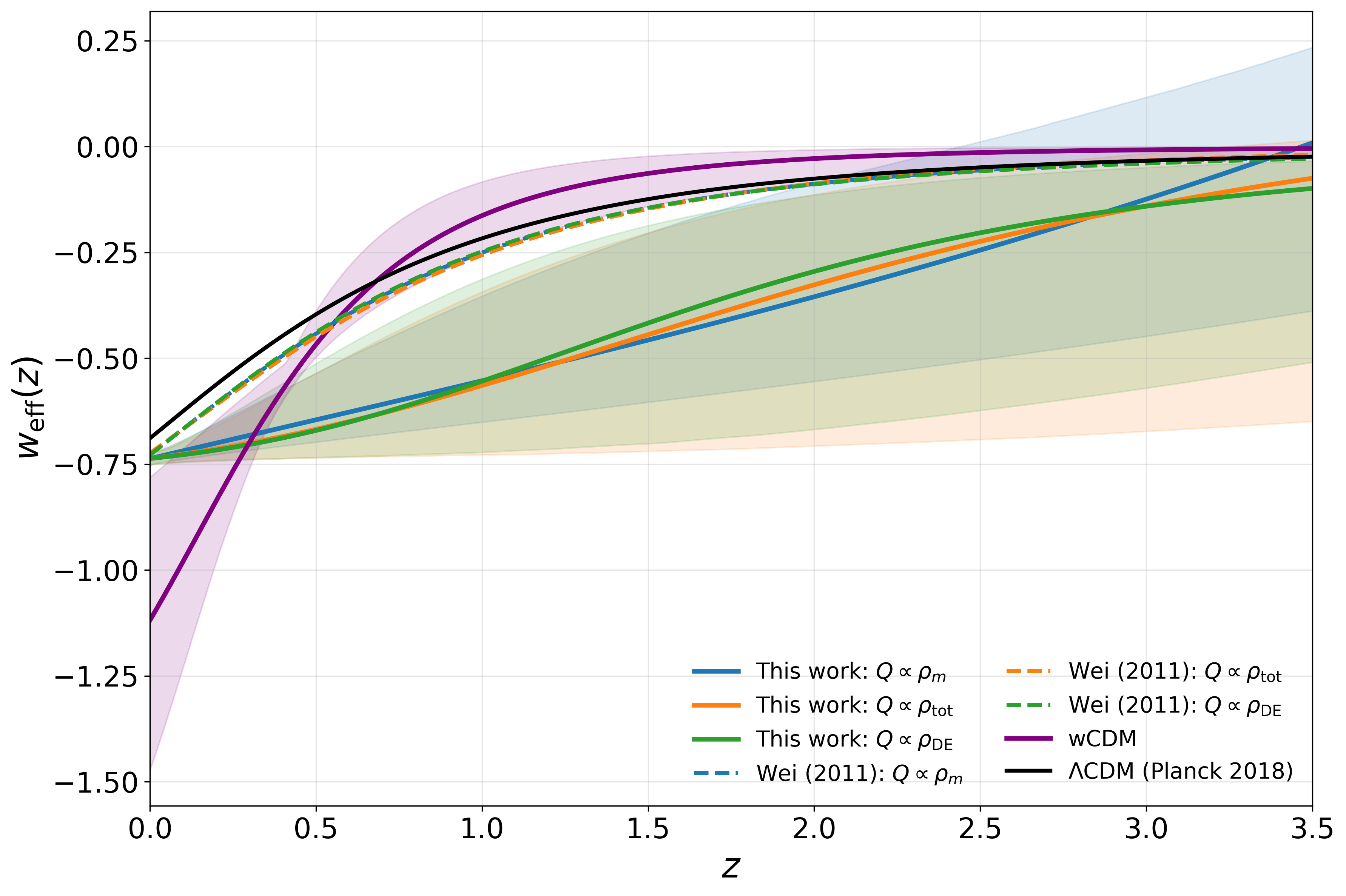}
\caption{Evolution of the effective equation of state $w_{\mathrm{eff}}(z)$ as a function of redshift for the three interacting dark energy models considered in this work: $Q \propto \rho_{dm}$, $Q \propto \rho_{\rm tot}$, and $Q \propto \rho_{\rm DE}$. Solid curves show the posterior median reconstructions obtained from the strong-lensing joint constraints, while shaded regions represent the $3\sigma$ confidence regions. Dashed curves correspond to reconstruction from the constraints reported by \citet{wei_interaction_2011}, obtained using SNe~Ia, BAO, and CMB data. The purple solid curve represents the reconstructed $w$CDM model, and the black curve indicates the $\Lambda$CDM prediction from Planck 2018.
}
\label{fig:Weff}
\end{figure}

\subsection{Comparison of the effective equation of state across models}

Figure~\ref{fig:Weff} shows the reconstructed evolution of $w_{\mathrm{eff}}(z)$
for the three interacting models considered in this work, the
$\Lambda$CDM prediction from Planck~2018, the $w$CDM scenario, and the results reported by \citet{wei_interaction_2011}.
In all interacting cases, the effective equation of state departs from the $\Lambda$CDM behavior, exhibiting less negative values at low redshifts and a smoother transition toward $w_{\mathrm{eff}} \simeq 0$ at earlier times.

Although the $w_{\mathrm{eff}}(z)$ reconstruction presented by \citet{wei_interaction_2011} shows small departures from $\Lambda$CDM at low redshift, our constraints on interacting parameters exhibit larger departures within the $2\sigma$ confidence region from $\Lambda$CDM. At higher redshift, the model of \citet{wei_interaction_2011} turns out to be closer to $\Lambda$CDM than the models considered here. This behavior reflects the sensitivity of lensing observables to modifications of the expansion history. Our results show that the differences between the models become more noticeable at intermediate redshifts, showing how the specific density entering the interacting term modulates the expansion history. This behavior emphasizes that strong lensing primarily constrains late-time cosmic dynamics.

Despite the different functional forms of the interaction term, all three interacting
models exhibit a remarkably similar value of the effective equation of state at the
present epoch, converging to $w_{\mathrm{eff}}(z=0)\simeq -0.7$. This behavior reflects the
fact that strong gravitational lensing tightly constrains the late-time expansion rate, yielding the total effective cosmic fluid to reproduce a comparable present-day
acceleration, independently of the $Q$ interacting term. In interacting
scenarios, this requirement is satisfied through a redistribution of energy between
dark matter and dark energy, rather than by tuning a single equation-of-state parameter.

This behavior contrasts  with the $w$CDM case constrained by the same lensing
data, where the absence of dark-sector interactions leaves the equation-of-state
parameter $w$ as the only degree of freedom. As a result, the lensing constraints drive the $w$CDM model toward
$w\simeq -1$, effectively reproducing a $\Lambda$CDM-like behavior. 

These different dynamics suggest that $w_{\mathrm{eff}}$ is a diagnostic tool to distinguish between dark interaction strength at low and intermediate redshifts

\subsection{Dynamics of the interaction strength}
\label{sec:interaction_dynamics}

\begin{figure}[htbp]
\centering
\includegraphics[width=0.45\textwidth]{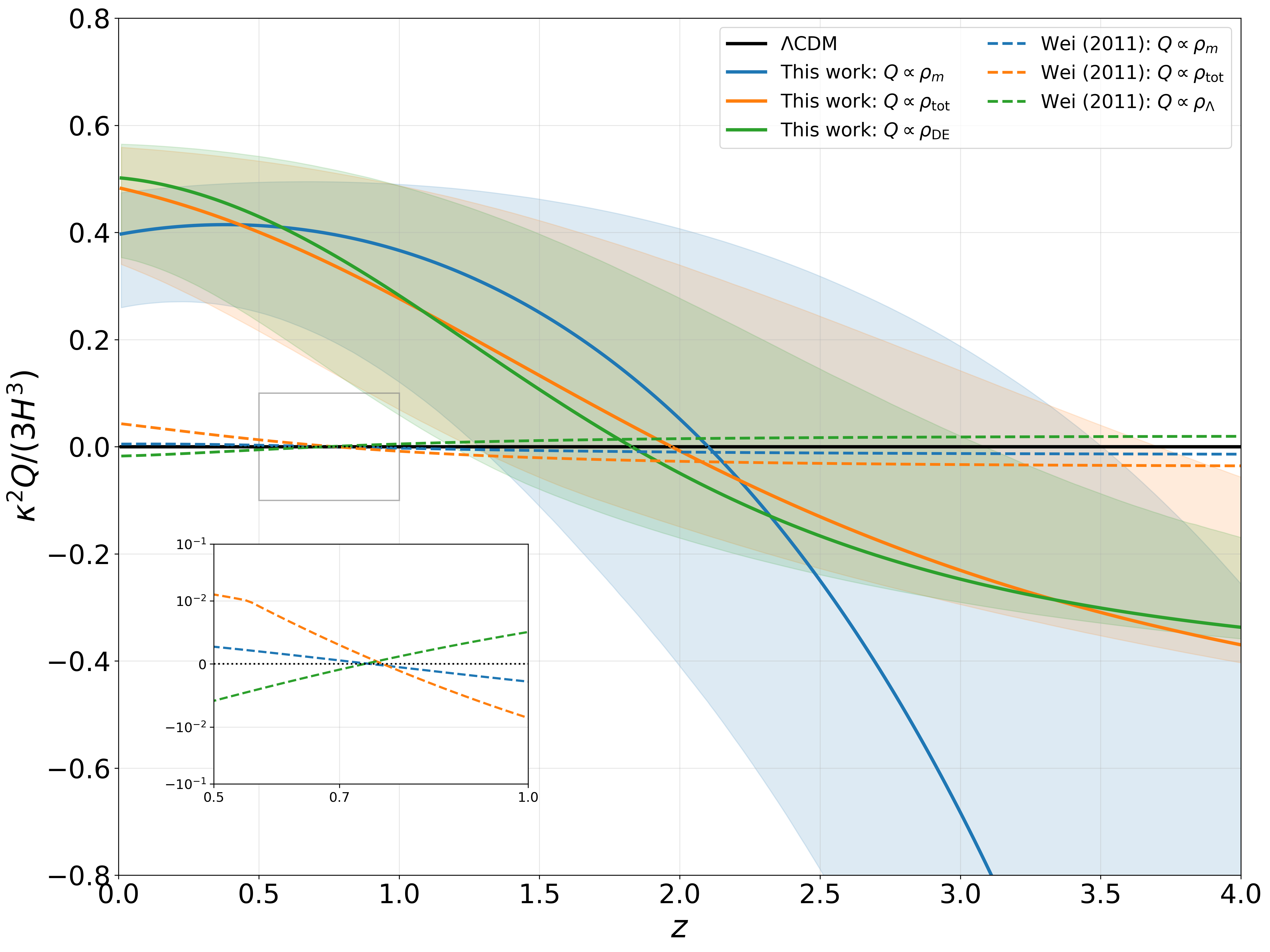}
\caption{Evolution of the dimensionless interaction parameter $\kappa^2 Q / 3H^3$ as a function of redshift for the three interacting dark energy models considered in this work: $Q \propto \rho_{dm}$, $Q \propto \rho_{\rm tot}$, and $Q \propto \rho_{\rm DE}$. Solid curves show the posterior median reconstructions obtained from the strong-lensing sample, while shaded regions represent the $2\sigma$ confidence regions. Dashed curves correspond to the constraints reported by \citet{wei_interaction_2011}, derived from SNe~Ia, BAO, and CMB data. The horizontal line at zero marks the $\Lambda$CDM limit ($Q=0$).}
\label{fig:Qz}
\end{figure}

The dimensionless quantity $\kappa^2 Q / 3H^3$ provides a physical measure of the interaction strength, normalized to the expansion rate of the Universe. Unlike the coupling parameter $\beta$ alone, this quantity allows a direct comparison of the effective energy transfer across different interaction models and cosmic epochs. The joint reconstruction shown in Fig.~\ref{fig:Qz} highlights clear and systematic differences between the three interacting scenarios.

In all models, the present-day value of
$\kappa^2 Q / 3H^3$ is significantly different from zero, indicating that
the interaction inferred from strong gravitational lensing is not a
negligible correction to the background dynamics. In contrast with
previous analyzes using SNe~Ia, BAO, and CMB data \citep{wei_interaction_2011}, where the interaction amplitude
remains close to zero, the lensing-based constraints favor values that
are dynamically relevant when normalized to the Hubble expansion rate.

At $z=0$, the interaction strength reaches values of order $\kappa^2 Q / 3H^3 \simeq 0.4$ for $Q \propto \rho_{dm}$, while both $Q \propto \rho_{\rm tot}$ and $Q \propto \rho_{\rm DE}$ attain slightly larger amplitudes, with present-day values close to $\simeq 0.5$. This pattern reflects the different density weights entering the interaction term and illustrates how the same sign-changeable mechanism can lead to distinct effective couplings at the present epoch. For the $Q \propto \rho_{\rm DE}$ case, the interaction amplitude
is modulated by the evolution of the dark energy density parameter, which enhances its impact at late times when $\Omega_{\rm DE}$ dominates. However, because the normalized quantity $\kappa^2 Q / 3H^3$
depends simultaneously on $\beta$ and $q(z)$, the three models yield comparable low-redshift amplitudes within the reconstructed confidence regions.

Fig.~\ref{fig:Qz} illustrates that strong gravitational lensing
is sensitive to both the sign and the normalized dynamical impact of dark-sector interactions. The comparison with SNe~Ia, BAO, and CMB based results highlights that lensing data are a  complementary probe in
redshift regimes where the effective coupling can be dynamically significant \citep{2004PhRvD..70d3534L,2011PhRvD..84l3529L}.

\subsection{Comparison of interacting models}

Table~\ref{comparison_table} summarizes the statistical comparison between the
interacting dark energy models and the $w$CDM scenario using the AIC and BIC information criteria. Both indicators consistently favor models with a sign-changeable interaction over the $w$CDM case, while revealing a clear statistical hierarchy among the interacting models.

According to both AIC and BIC, the model with an interaction proportional to the
dark energy density ($Q \propto \rho_{\rm DE}$) provides the best statistical
performance, followed by the model coupled to the total density ($Q \propto \rho_{\rm tot}$), and finally the interaction proportional to the dark matter density
($Q \propto \rho_{dm}$). The $w$CDM model is strongly disfavored
with respect to all interacting scenarios, with $\Delta$AIC and $\Delta$BIC larger than 8, indicating strong evidence against this scenario.

Although the $Q \propto \rho_{\rm DE}$ model, with $\Omega_{dm0} = 0.2634^{+0.0051}_{-0.0052}$,  is statistically preferred, it is
worth noting that information criteria quantify only the balance between
goodness of fit and model complexity. From a phenomenological perspective, the
interaction proportional to the total density $Q \propto \rho_{tot}$, represents an intermediate scenario, avoiding the extremes of coupling exclusively to dark matter or dark
energy, and yielding a moderate interaction strength ($\beta = -0.2662^{+0.0343}_{-0.0253}$).

A comparison with the results of \citet{wei_interaction_2011} reveals a 
different ranking of the interaction models. Their analysis favors the non-interacting $\Lambda$CDM
scenario, while the interacting models exhibit a weak coupling
$|\beta|\sim 10^{-2}$. In contrast, the strong-lensing constraints employed in
this work favor interacting scenarios with significantly larger
interaction strengths. In addition, it is important to note that in the four models discussed here, $\Omega_{dm0}$ is  slightly lower than $\Omega_{dm0} = 0.265 \pm 0.005$ reported by \cite{Planck_2018}.

The above discrepancies suggest that different cosmological probes may be sensitive
to distinct aspects of interacting dark energy models. Strong gravitational lensing probes the cosmic geometry through distance ratios that are sensitive to the integrated expansion history \citep{2012JCAP...03..016C}, offering a complementary sensitivity to dynamical energy exchange in the dark sector, since gravitational lensing provides complementary constraints on dark energy parameters relative to other probes \citep{2015ApJ...813...69M}. On the other hand,  systematics related to
lens modeling and mass-profile assumptions cannot be fully excluded
\citep{Magana2018ApJ, verdugo_EPJC_2024}.

\section{Conclusions}\label{sec:Conclusions}

In this work, we have used strong gravitational lensing at galactic and cluster
scales to constrain three sign-changeable interacting dark energy models in which the coupling term depends explicitly on the deceleration parameter and is proportional to different dark-sector densities. By combining a sample of early-type galaxy lenses with the well-studied cluster Abell 1689, we provided joint constraints on three interaction models:
$Q \propto \rho_{dm}$, $Q \propto \rho_{\mathrm{tot}}$, and $Q \propto \rho_{\mathrm{DE}}$. For $Q$ coupled with dark matter density, we found $\Omega_{dm0} = 0.2631^{+0.0050}_{-0.0051}$ and $\beta = -0.8290^{+0.1334}_{-0.1053}$. The case proportional to the total density yields $\Omega_{dm0} = 0.2632^{+0.0049}_{-0.0049}$ and $\beta = -0.2662^{+0.0343}_{-0.0253}$, whereas the dark energy-dependent model results in $\Omega_{dm0} = 0.2634^{+0.0051}_{-0.0052}$ and $\beta = -0.3756^{+0.0430}_{-0.0302}$.

All interacting scenarios consistently favor dark matter density parameters
$\Omega_{dm0}$ consistent within $1\sigma$ with the Planck 2018 estimate  ($\Omega_{dm0} = 0.265 \pm 0.005$). Furthermore, the $\beta$ interaction parameters are negative and
substantially larger than those obtained from other probes such as
SNe~Ia, BAO, and CMB \citep{wei_interaction_2011}, indicating an efficient transfer of energy from dark
energy to dark matter during the late-time accelerated phase. A direct
consequence of this interaction is an early transition to cosmic acceleration,
with transition redshifts $z_t \sim 1.7$-$1.9$, larger than the $\Lambda$CDM expectation. Despite these deviations from the standard model, the reconstructed expansion histories remain consistent with cosmic chronometer measurements within $2\sigma$ confidence regions.

The analysis of $w_{\mathrm{eff}}(z)$ reveals
that, although the interacting models differ significantly in their functional form, they converge toward a similar present day value,
$w_{\mathrm{eff}}(z=0) \simeq -0.7$. This convergence reflects the fact that
strong lensing robustly fixes the current acceleration rate of the Universe,
forcing different interacting scenarios to reproduce comparable late time
kinematics through energy exchange between dark components. In contrast, a
non interacting $w$CDM model constrained by the same data is driven toward
$w \simeq -1$, effectively mimicking $\Lambda$CDM behavior. Although the specific coupling functions and datasets differ, our results are consistent with the general picture presented in \citet{2017JCAP...01..028C}, where a coupling between dark matter and dark energy modifies the effective expansion dynamics and the inferred dark energy behavior.

The dimensionless interaction measure $\kappa^2 Q / 3H^3$ further shows that strong gravitational lensing is sensitive not only to the sign of the interaction, but also to its normalized dynamical impact. In all interacting models, the present day interaction strength is significantly different from zero, in contrast to previous results by \citet{wei_interaction_2011}. This reinforces the role of strong lensing as a complementary probe of interacting dark-sector scenarios, whose observational constraints are known to depend on the coupling prescription and dataset combination \citep{Wang2016}. Recent analyses based on Fermi GRBs and DESI DR2 data find no significant preference for interacting dark energy over $\Lambda$CDM according to AIC and BIC criteria \citep{2026JHEAp..5100534Z}, while constraints using CMB, DESI DR2, SNe~Ia, and cosmic chronometers suggest at most mild evidence for an interaction, with model-selection criteria giving a mixed picture \citep{2026PhRvD.113b3515P}. In contrast, our results indicate relevant interaction strengths when constrained with strong-lensing observables. This result highlights the complementary nature of strong lensing as a probe of dark sector physics  \citep{verdugo_EPJC_2024,Magana2018ApJ}. Additionally, we found that the model $Q\propto \rho_{\mathrm{DE}}$ provides the best overall performance. However, from a physical standpoint, the model $Q \propto \rho_{\mathrm{tot}}$ emerges as a  balanced intermediate scenario, producing a dynamically significant effect on the expansion history.

Our findings demonstrate that strong gravitational lensing provides a powerful and independent toll to test interacting dark energy scenarios and to explore the physics of the dark sector beyond the standard $\Lambda$CDM paradigm. In this context, our results support the broader view that increasingly accurate and diverse cosmological probes are essential to determine whether current tensions point to new  physics in the dark sector or instead reflect unknown systematic effects \citep{2024RPPh...87c6901W}.

\begin{table*}[t] 
\centering
\caption{Statistical comparison of interacting dark energy cosmological models. Parameter values correspond to posterior medians with 68\% credible intervals. The goodness of fit and information criteria (AIC and BIC) are evaluated at the posterior median of the model parameters.}
\label{comparison_table}

\vspace{0.3cm} 
\small 
\setlength{\tabcolsep}{10pt} 
\renewcommand{\arraystretch}{1.6} 

\begin{tabular*}{\textwidth}{@{\extracolsep{\fill}}lcccc}
\hline\hline
Model 
& $Q = 3\beta qH\rho_{DE}$
& $Q = 3\beta qH\rho_{dm}$
& $Q = 3\beta qH\rho_{\mathrm{tot}}$
& $w$CDM \\
\hline

Median
& $\begin{aligned}
\Omega_{dm0} &= 0.2634^{+0.0051}_{-0.0052} \\
\beta &= -0.3756^{+0.0430}_{-0.0302}
\end{aligned}$
& $\begin{aligned}
\Omega_{dm0} &= 0.2631^{+0.0050}_{-0.0051} \\
\beta &= -0.8290^{+0.1334}_{-0.1053}
\end{aligned}$
& $\begin{aligned}
\Omega_{dm0} &= 0.2632^{+0.0049}_{-0.0049} \\
\beta &= -0.2662^{+0.0343}_{-0.0253}
\end{aligned}$
& $\begin{aligned}
\Omega_{dm0} &= 0.2621^{+0.0051}_{-0.0050} \\
w &= -1.5156^{+0.1834}_{-0.2110}
\end{aligned}$ \\

\hline

$\chi^2_{med}$
& 283.08
& 294.34
& 284.45
& 299.90 \\

$\chi^2_{med}/\mathrm{dof}$
& 1.887
& 1.962
& 1.896
& 1.999 \\

AIC
& 325.08
& 336.34
& 326.45
& 341.90 \\

BIC
& 391.06
& 402.32
& 392.43
& 407.87 \\

$\Delta$AIC
& 0.00
& 11.26
& 1.37
& 16.81 \\

$\Delta$BIC
& 0.00
& 11.26
& 1.37
& 16.81 \\

\hline
Rank & \textbf{1} & 3 & 2 & 4 \\
\hline\hline
\end{tabular*}

\vspace{0.1cm}
\end{table*}

\appendix
\section{Parameter regions leading to non-physical evolution}\label{App1}

In this appendix we clarify the parameter regions where the background evolution becomes non-physical because
the reconstructed densities lead to $H^2(z)<0$. In order to identify the regions where the background evolution becomes non-physical,
it is convenient to isolate the analytical factor that controls the sign of the effective
densities entering the Friedmann equation. We therefore define
\begin{equation}
F(z) \equiv \frac{N(z)}{D},
\end{equation}
where $N(z)$ is a redshift-dependent function and $D>0$ is a constant denominator.
Depending on the interaction model, $F(z)$ corresponds to $B(z)$ for $Q\propto\rho_{\rm DE}$
or to $e(z)$ for $Q\propto\rho_{\rm tot}$. In all cases, $F(z)$ appears multiplicatively in the expression of $E^2(z)$, so that
a negative value of $F(z)$ can drive the total expansion rate outside the physical domain.

The numerical computation of $H(z)$  relies on the Friedmann equation,
\begin{equation}
H^2(z)=H_0^2\,E^2(z),
\end{equation}
with $E^2(z)$ built from the (effective) densities of each sector.
The equation is only well-defined when
\begin{equation}
E^2(z)\ge 0 \qquad \text{for all redshifts used in the likelihood.}
\end{equation}
In some regions of parameter space (typically for sufficiently negative $\beta$)
the analytical expressions can yield an effective density that becomes negative over a finite
redshift interval. When that drives $E^2(z)<0$, $H^2(z)=H_0^2E^2(z)$ becomes negative, so
$H(z)$ becomes imaginary and the model must be discarded as non-physical
in the context of a real FRW expansion history.

\subsection{Case: $Q \propto \rho_{\rm DE}$}
\label{app:case3}

\begin{figure}[!ht]
\centering
\includegraphics[width=0.45\textwidth]{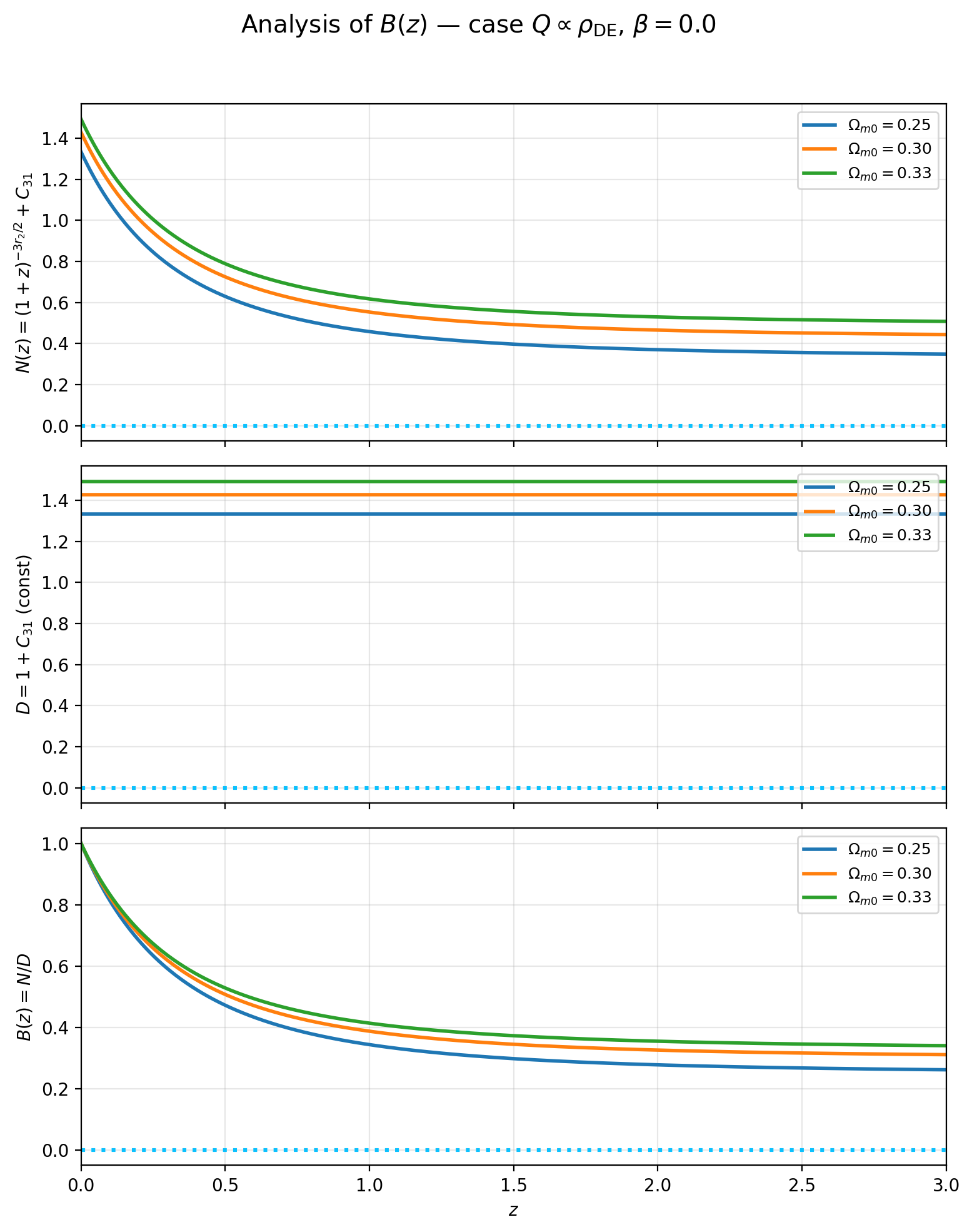}
\caption{Case $Q\propto \rho_{\rm DE}$.  Numerator $N(z)$, constant denominator $D$, and ratio
$B(z)=N(z)/D$ for representative values of $\Omega_{dm0}$. With  $\beta=0$ the figure illustrates that the
analytical factor remains strictly positive.}
\label{fig:app_case3_Bz}
\end{figure}

\begin{figure}[!ht]
\centering
\includegraphics[width=0.45\textwidth]{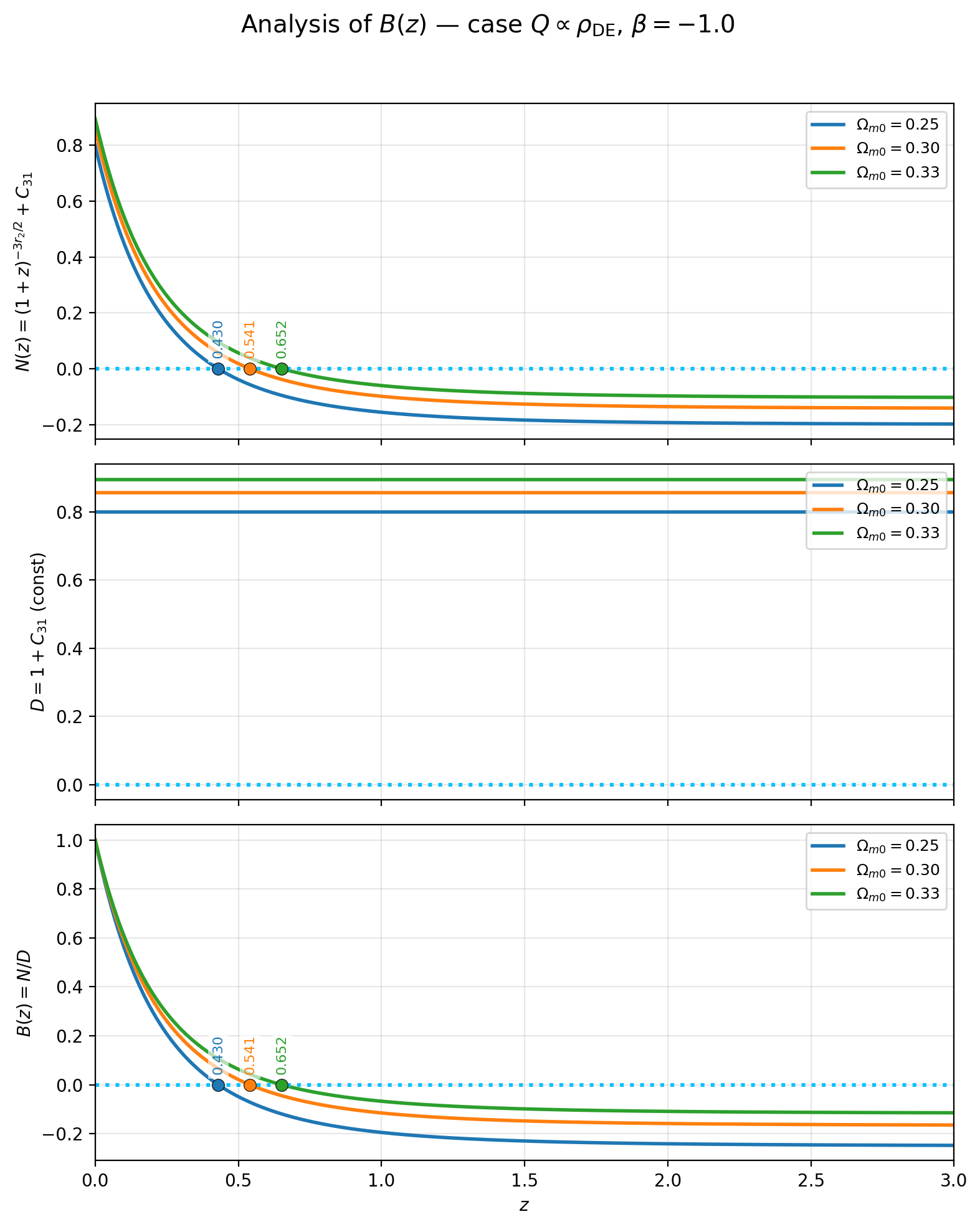}
\caption{Numerator $N(z)$, constant denominator $D$, and ratio
$B(z)=N(z)/D$ for representative values of $\Omega_{dm0}$. With   $\beta<0$ panel shows that $N(z)$ can cross
zero at $z=z_\star$, inducing a sign change in $e(z)$ and potentially pushing the background outside
the real domain where $H^2(z)\ge 0$.}
\label{fig:app_case4_Bz}
\end{figure}

Figures~\ref{fig:app_case3_Bz} and ~\ref{fig:app_case4_Bz} show the behavior of the numerator $N(z)$, the denominator $D$,
and their ratio $B(z)=N(z)/D$ for the interaction proportional to the dark energy density,
for representative values of $\Omega_{dm0}$ and two choices of $\beta$.

For $\beta=0$, the numerator remains positive and monotonically decreases with $z$, while the
denominator is strictly positive and constant; consequently $B(z)>0$ over the full range and no additional sign flip is induced by the
analytical prefactor entering the Friedmann equation.

For $\beta<0$, the same monotonic decrease of $N(z)$ can lead to a zero crossing at
$z=z_\star(\Omega_{dm0})$, which implies a sign flip in $B(z)$ and therefore modifies the sign
structure of the interaction term. Importantly, the same parameter choices may also render the
effective background densities non-physical (i.e. produce $E^2(z)<0$ over some interval), which is
the direct reason why $H(z)$ cannot be evaluated as a real function for those regions.

\subsection{Case: $Q \propto \rho_{\rm tot}$}
\label{app:case4}

An analogous diagnostic holds for the interaction proportional to the total density.
Figure~\ref{fig:app_case4_Bz} shows the corresponding $N(z)$, $D$, and $e(z)$ behaviour.
As in the other case, sufficiently negative $\beta$ can be associated with a zero crossing in $N(z)$,
together with a loss of physicality when the reconstructed background violates $E^2(z)\ge 0$.

\begin{figure}[!ht]
\centering
\includegraphics[width=0.45\textwidth]{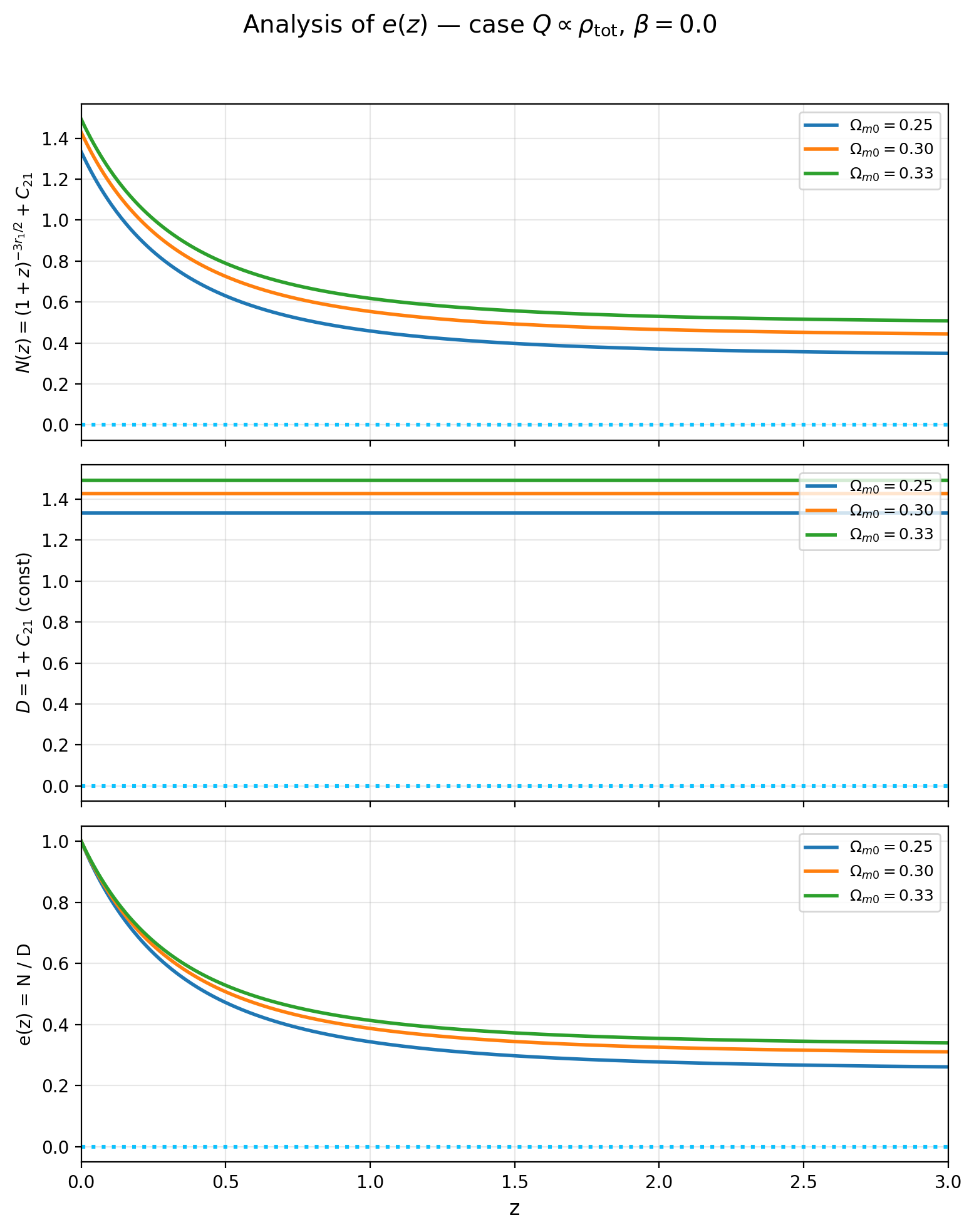}
\caption{Case $Q\propto \rho_{\rm tot}$. Numerator $N(z)$, constant denominator $D$, and ratio
$e(z)=N(z)/D$ for representative values of $\Omega_{dm0}$. With  $\beta=0$ the figure illustrates that the
analytical factor remains strictly positive.}
\label{fig:Erhotot_components_optionb}
\end{figure}

\begin{figure}[!ht]
\centering
\includegraphics[width=0.45\textwidth]{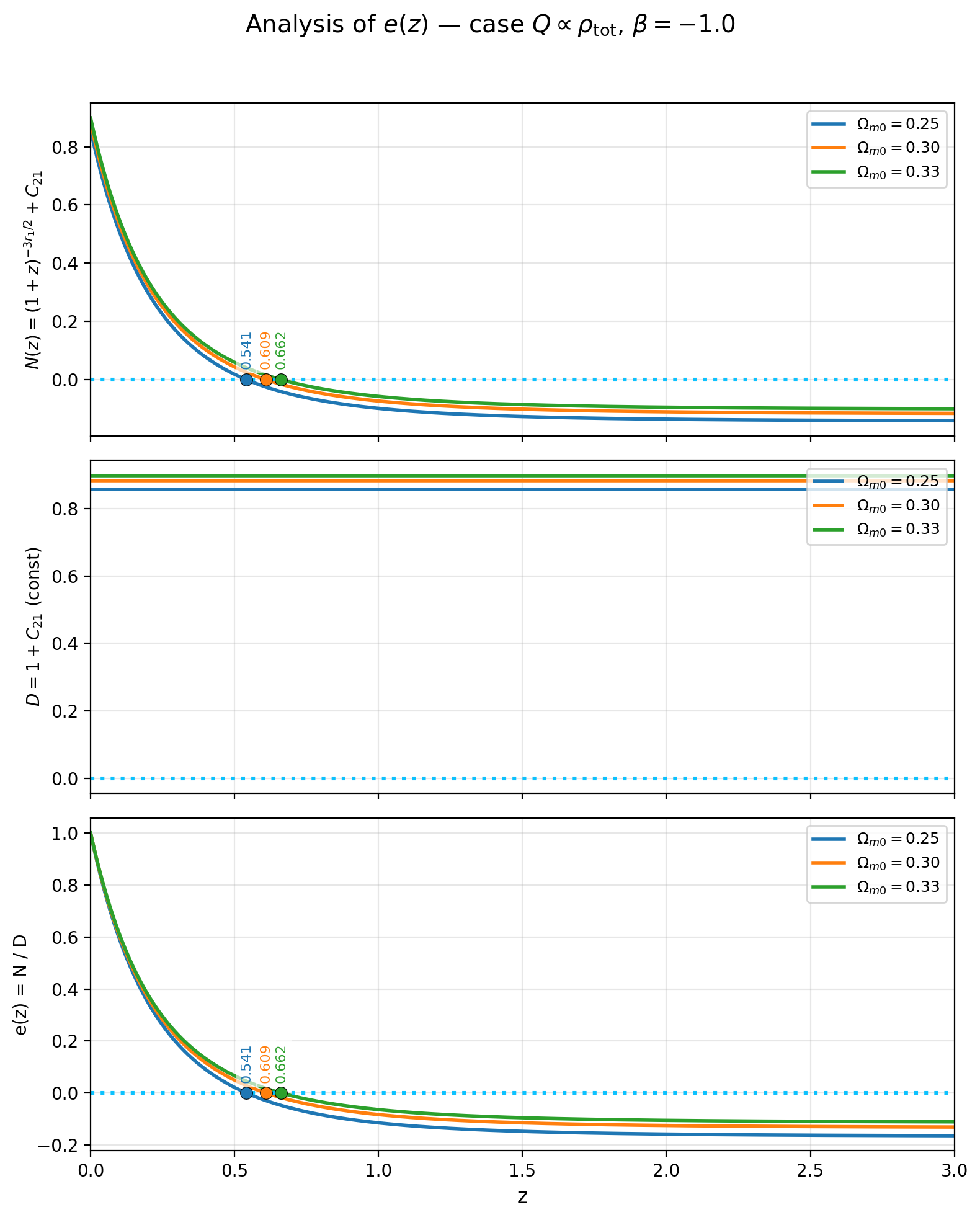}
\caption{Case $Q\propto \rho_{\rm tot}$. Numerator $N(z)$, constant denominator $D$, and ratio
$e(z)=N(z)/D$ for representative values of $\Omega_{dm0}$. With   $\beta<0$ panel shows that $N(z)$ can cross
zero at $z=z_\star$, inducing a sign change in $e(z)$ and potentially pushing the background outside
the real domain where $H^2(z)\ge 0$.}
\label{fig:Erhotot_components_optionA_labels_above_points}
\end{figure}



\section*{Acknowledgements}
F. Villalobos, T. Verdugo, J. Magaña acknowledge support from the Secretaría de Ciencia, Humanidades, Tecnología e Innovación (SECIHTI) under grant CBF-2025-I-551 (“Convocatoria Ciencia Básica y de Frontera 2025”).
FV, JM, PTI, TV acknowledge the support provided by the Agencia Nacional de Investigación y Desarrollo (ANID) through its program Fomento a la Vinculación Internacional (FOVI) 240098. FV and JM acknowledge and the Fondo Nacional de Desarrollo Científico y Tecnológico Fondecyt Regular No. 1240514.
FV acknowledges the use supercomputer of GÜINA (EQM 200216) and support from Universidad Central de Chile from "Facultad de Ingeniería y Arquitectura", "Vicerectoría de Investigación", and the project "Puente de Investigación UCEN 2025".

\bibliographystyle{elsarticle-harv} 
\bibliography{bibliografia.bib}

@article{Caminha:2021iwo,
    author = "Caminha, G. B. and Suyu, S. H. and Grillo, C. and Rosati, P.",
    title = "{Galaxy cluster strong lensing cosmography - Cosmological constraints from a sample of regular galaxy clusters}",
    eprint = "2110.06232",
    archivePrefix = "arXiv",
    primaryClass = "astro-ph.CO",
    doi = "10.1051/0004-6361/202141994",
    journal = "Astron. Astrophys.",
    volume = "657",
    pages = "A83",
    year = "2022"
}

@article{Carrasco:2023imi,
    author = "Carrasco, Raul and Rincon, Angel and Saavedra, Joel and Videla, Nelson",
    title = "{Discriminating interacting dark energy models using Statefinder diagnostic}",
    eprint = "2310.04324",
    archivePrefix = "arXiv",
    primaryClass = "gr-qc",
    doi = "10.1140/epjc/s10052-024-12733-6",
    journal = "Eur. Phys. J. C",
    volume = "84",
    number = "5",
    pages = "459",
    year = "2024"
}

@article{Arevalo:2019axj,
    author = "Arevalo, Fabiola and Cid, Antonella and Chimento, Luis P. and Mella, Patricio",
    title = "{On sign-changeable interaction in FLRW cosmology}",
    eprint = "1901.04300",
    archivePrefix = "arXiv",
    primaryClass = "gr-qc",
    doi = "10.1140/epjc/s10052-019-6872-7",
    journal = "Eur. Phys. J. C",
    volume = "79",
    number = "4",
    pages = "355",
    year = "2019"
}

@article{CosmoVerseNetwork:2025alb,
    author = "Di Valentino, Eleonora and others",
    collaboration = "CosmoVerse Network",
    title = "{The CosmoVerse White Paper: Addressing observational tensions in cosmology with systematics and fundamental physics}",
    eprint = "2504.01669",
    archivePrefix = "arXiv",
    primaryClass = "astro-ph.CO",
    doi = "10.1016/j.dark.2025.101965",
    journal = "Phys. Dark Univ.",
    volume = "49",
    pages = "101965",
    year = "2025"
}

@article{Bolotin:2013jpa,
    author = "Bolotin, Yu. L. and Kostenko, A. and Lemets, O. A. and Yerokhin, D. A.",
    title = "{Cosmological Evolution With Interaction Between Dark Energy And Dark Matter}",
    eprint = "1310.0085",
    archivePrefix = "arXiv",
    primaryClass = "astro-ph.CO",
    doi = "10.1142/S0218271815300074",
    journal = "Int. J. Mod. Phys. D",
    volume = "24",
    number = "03",
    pages = "1530007",
    year = "2014"
}

@ARTICLE{diValentino:2017,
       author = {{Di Valentino}, Eleonora and {Melchiorri}, Alessandro and {Mena}, Olga},
        title = "{Can interacting dark energy solve the H$_{0}$ tension?}",
      journal = {Physical Review D},
         year = 2017,
        month = aug,
       volume = {96},
       number = {4},
          eid = {043503},
        pages = {043503},
          doi = {10.1103/PhysRevD.96.043503},
archivePrefix = {arXiv},
       eprint = {1704.08342},
 primaryClass = {astro-ph.CO},
       adsurl = {https://ui.adsabs.harvard.edu/abs/2017PhRvD..96d3503D}
}

@article{Motta:2021hvl,
    author = "Motta, Ver{\'o}nica and Garc{\'\i}a-Aspeitia, Miguel A. and Hern{\'a}ndez-Almada, Alberto and Maga{\~n}a, Juan and Verdugo, Tom{\'a}s",
    title = "{Taxonomy of Dark Energy Models}",
    eprint = "2104.04642",
    archivePrefix = "arXiv",
    primaryClass = "astro-ph.CO",
    doi = "10.3390/universe7060163",
    journal = "Universe",
    volume = "7",
    number = "6",
    pages = "163",
    year = "2021"
}

@article{Scolnic:2021amr,
    author = "Scolnic, Dan and others",
    title = "{The Pantheon+ Analysis: The Full Data Set and Light-curve Release}",
    eprint = "2112.03863",
    archivePrefix = "arXiv",
    primaryClass = "astro-ph.CO",
    doi = "10.3847/1538-4357/ac8b7a",
    journal = "Astrophys. J.",
    volume = "938",
    number = "2",
    pages = "113",
    year = "2022"
}

@ARTICLE{Riess:2022ApJ,
       author = {{Riess}, Adam G. and {Yuan}, Wenlong and {Macri}, Lucas M. and {Scolnic}, Dan and {Brout}, Dillon and {Casertano}, Stefano and {Jones}, David O. and {Murakami}, Yukei and {Anand}, Gagandeep S. and {Breuval}, Louise and {Brink}, Thomas G. and {Filippenko}, Alexei V. and {Hoffmann}, Samantha and {Jha}, Saurabh W. and {D'arcy Kenworthy}, W. and {Mackenty}, John and {Stahl}, Benjamin E. and {Zheng}, WeiKang},
        title = "{A Comprehensive Measurement of the Local Value of the Hubble Constant with 1 km s$^{-1}$ Mpc$^{-1}$ Uncertainty from the Hubble Space Telescope and the SH0ES Team}",
      journal = {The Astrophysical Journall},
         year = 2022,
        month = jul,
       volume = {934},
       number = {1},
          eid = {L7},
        pages = {L7},
          doi = {10.3847/2041-8213/ac5c5b},
archivePrefix = {arXiv},
       eprint = {2112.04510},
 primaryClass = {astro-ph.CO},
       adsurl = {https://ui.adsabs.harvard.edu/abs/2022ApJ...934L...7R}
}

@ARTICLE{riess_SN1a_1998,
   author = {{Riess}, A.~G. and {Filippenko}, A.~V. and {Challis}, P. and 
        {Clocchiatti}, A. and {Diercks}, A. and {Garnavich}, P.~M. and
        {Gilliland}, R.~L. and {Hogan}, C.~J. and {Jha}, S. and 
        {Kirshner}, R.~P. and {Leibundgut}, B. and {Phillips}, M.~M. and
        {Reiss}, D. and {Schmidt}, B.~P. and {Schommer}, R.~A. and
        {Smith}, R.~C. and {Spyromilio}, J. and {Stubbs}, C. and
        {Suntzeff}, N.~B. and {Tonry}, J.},
    title = "{Observational Evidence from Supernovae for an Accelerating Universe and a Cosmological Constant}",
  journal = {The Astronomical Journal},
     year = 1998,
   volume = 116,
    pages = {1009-1038},
      doi = {10.1086/300499},
   adsurl = {https://ui.adsabs.harvard.edu/abs/1998AJ....116.1009R}
}

@ARTICLE{perlmutter_SN1a_1999,
   author = {{Perlmutter}, S. and {Aldering}, G. and {Goldhaber}, G. and
        {Knop}, R.~A. and {Nugent}, P. and {Castro}, P.~G. and
        {Deustua}, S. and {Fabbro}, S. and {Goobar}, A. and
        {Groom}, D.~E. and {Hook}, I.~M. and {Kim}, A.~G. and
        {Kim}, M.~Y. and {Lee}, J.~C. and {Nunes}, N.~J. and
        {Pain}, R. and {Pennypacker}, C.~R. and {Quimby}, R. and
        {Lidman}, C. and {Ellis}, R.~S. and {Irwin}, M. and
        {McMahon}, R.~G. and {Ruiz-Lapuente}, P. and {Walton}, N. and
        {Schaefer}, B. and {Boyle}, B.~J. and {Filippenko}, A.~V. and
        {Matheson}, T. and {Fruchter}, A.~S. and {Panagia}, N. and
        {Newberg}, H.~J.~M. and {Couch}, W.~J. and 
        {The Supernova Cosmology Project}},
    title = "{Measurements of {$\Omega$} and {$\Lambda$} from 42 High-Redshift Supernovae}",
  journal = {The Astrophysical Journal},
     year = 1999,
   volume = 517,
    pages = {565-586},
      doi = {10.1086/307221},
   adsurl = {https://ui.adsabs.harvard.edu/abs/1999ApJ...517..565P}
}

@ARTICLE{Planck_2018,
   author = {{Planck Collaboration} and {Aghanim}, N. and {Akrami}, Y. and 
        {Ashdown}, M. and {Aumont}, J. and {Baccigalupi}, C. and 
        {Ballardini}, M. and {Banday}, A.~J. and {Barreiro}, R.~B. and 
        {Bartolo}, N. and {others}},
    title = "{Planck 2018 results. VI. Cosmological parameters}",
  journal = {Astronomy and Astrophysics},
     year = 2020,
   volume = 641,
    pages = {A6},
      doi = {10.1051/0004-6361/201833910},
   adsurl = {https://ui.adsabs.harvard.edu/abs/2020A&A...641A...6P}
}

@ARTICLE{wei_interaction_2011,
   title={Cosmological Constraints on the Sign-Changeable Interactions},
   volume={56},
   ISSN={0253-6102},
   url={http://dx.doi.org/10.1088/0253-6102/56/5/29},
   DOI={10.1088/0253-6102/56/5/29},
   number={5},
   journal={Communications in Theoretical Physics},
   publisher={IOP Publishing},
   author={Wei, Hao},
   year={2011},
   month=nov, pages={972–980} }

@article{amante_SL_2020,
   title={Testing dark energy models with a new sample of strong-lensing systems},
   volume={498},
   ISSN={1365-2966},
   url={http://dx.doi.org/10.1093/mnras/staa2760},
   DOI={10.1093/mnras/staa2760},
   number={4},
   journal={Monthly Notices of the Royal Astronomical Society},
   publisher={Oxford University Press (OUP)},
   author={Amante, Mario H and Magaña, Juan and Motta, V and García-Aspeitia, Miguel A and Verdugo, Tomás},
   year={2020},
   month=sep, pages={6013–6033} }

@ARTICLE{jullo_SL_2010,
   author = {{Jullo}, E. and {Natarajan}, P. and {Kneib}, J.-P. and 
        {D'Aloisio}, A. and {Limousin}, M. and {Richard}, J. and {Schimd}, C.},
    title = "{Cosmological Constraints from Strong Gravitational Lensing in Clusters of Galaxies}",
  journal = {Science},
     year = 2010,
   volume = 329,
    pages = {924-927},
      doi = {10.1126/science.1185759},
   adsurl = {https://ui.adsabs.harvard.edu/abs/2010Sci...329..924J}
}

@ARTICLE{verdugo_EPJC_2024,
       author = {{Verdugo}, Tom{\'a}s and {Amante}, Mario H. and {Maga{\~n}a}, Juan and {Garc{\'\i}a-Aspeitia}, Miguel A. and {Hern{\'a}ndez-Almada}, Alberto and {Motta}, Ver{\'o}nica},
        title = "{Synchronize your chrono-brane: testing a variable brane tension model with strong gravitational lensing}",
      journal = {European Physical Journal C},
         year = 2024,
        month = jan,
       volume = {84},
       number = {1},
          eid = {93},
        pages = {93},
          doi = {10.1140/epjc/s10052-024-12434-0},
archivePrefix = {arXiv},
       eprint = {2401.06376},
 primaryClass = {astro-ph.CO},
       adsurl = {https://ui.adsabs.harvard.edu/abs/2024EPJC...84...93V}
}

@ARTICLE{moresco_CC_2016,
   author = "Moresco, Michele and Pozzetti, Lucia and Cimatti, Andrea and Jimenez, Raul and Maraston, Claudia and Verde, Licia and Thomas, Daniel and Citro, Annalisa and Tojeiro, Rita and Wilkinson, David",
    title = "{A 6{\%} measurement of the Hubble parameter at $z\sim0.45$: direct evidence of the epoch of cosmic re-acceleration}",
    eprint = "1601.01701",
    archivePrefix = "arXiv",
    primaryClass = "astro-ph.CO",
    doi = "10.1088/1475-7516/2016/05/014",
    journal = "JCAP",
    volume = "05",
    pages = "014",
    year = "2016"
}

@ARTICLE{DiValentino2021,
       author = {{Di Valentino}, Eleonora and {Mena}, Olga and {Pan}, Supriya and {Visinelli}, Luca and {Yang}, Weiqiang and {Melchiorri}, Alessandro and {Mota}, David F. and {Riess}, Adam G. and {Silk}, Joseph},
        title = "{In the realm of the Hubble tension-a review of solutions}",
      journal = {Classical and Quantum Gravity},
         year = 2021,
        month = jul,
       volume = {38},
       number = {15},
          eid = {153001},
        pages = {153001},
          doi = {10.1088/1361-6382/ac086d},
archivePrefix = {arXiv},
       eprint = {2103.01183},
 primaryClass = {astro-ph.CO},
       adsurl = {https://ui.adsabs.harvard.edu/abs/2021CQGra..38o3001D}
}

@ARTICLE{Magana2018ApJ,
       author = {{Maga{\~n}a}, Juan and {Acebr{\'o}n}, Ana and {Motta}, Ver{\'o}nica and {Verdugo}, Tom{\'a}s and {Jullo}, Eric and {Limousin}, Marceau},
        title = "{Strong Lensing Modeling in Galaxy Clusters as a Promising Method to Test Cosmography. I. Parametric Dark Energy Models}",
      journal = {The Astrophysical Journal},
         year = 2018,
        month = oct,
       volume = {865},
       number = {2},
          eid = {122},
        pages = {122},
          doi = {10.3847/1538-4357/aada7d},
archivePrefix = {arXiv},
       eprint = {1711.00829},
 primaryClass = {astro-ph.CO},
       adsurl = {https://ui.adsabs.harvard.edu/abs/2018ApJ...865..122M}
}

@ARTICLE{Copeland2006,
       author = {{Copeland}, Edmund J. and {Sami}, M. and {Tsujikawa}, Shinji},
        title = "{Dynamics of Dark Energy}",
      journal = {International Journal of Modern Physics D},
         year = 2006,
        month = jan,
       volume = {15},
       number = {11},
        pages = {1753-1935},
          doi = {10.1142/S021827180600942X},
archivePrefix = {arXiv},
       eprint = {hep-th/0603057},
 primaryClass = {hep-th},
       adsurl = {https://ui.adsabs.harvard.edu/abs/2006IJMPD..15.1753C}
}

@ARTICLE{Bamba2012,
       author = {{Bamba}, Kazuharu and {Capozziello}, Salvatore and {Nojiri}, Shin'ichi and {Odintsov}, Sergei D.},
        title = "{Dark energy cosmology: the equivalent description via different theoretical models and cosmography tests}",
      journal = {Astrophysics and Space Science
},
         year = 2012,
        month = nov,
       volume = {342},
       number = {1},
        pages = {155-228},
          doi = {10.1007/s10509-012-1181-8},
archivePrefix = {arXiv},
       eprint = {1205.3421},
 primaryClass = {gr-qc},
       adsurl = {https://ui.adsabs.harvard.edu/abs/2012Ap&SS.342..155B}
}

@ARTICLE{Wang2016,
       author = {{Wang}, B. and {Abdalla}, E. and {Atrio-Barandela}, F. and {Pav{\'o}n}, D.},
        title = "{Dark matter and dark energy interactions: theoretical challenges, cosmological implications and observational signatures}",
      journal = {Reports on Progress in Physics},
         year = 2016,
        month = sep,
       volume = {79},
       number = {9},
          eid = {096901},
        pages = {096901},
          doi = {10.1088/0034-4885/79/9/096901},
archivePrefix = {arXiv},
       eprint = {1603.08299},
 primaryClass = {astro-ph.CO},
       adsurl = {https://ui.adsabs.harvard.edu/abs/2016RPPh...79i6901W}
}

@article{Cai_Su_2010,
    author = {Cai, Rong-Gen and Su, Qiping},
    title = {On the dark sector interactions},
    journal = {Phys. Rev. D},
    volume = {81},
    pages = {103514},
    year = {2010},
    doi = {10.1103/PhysRevD.81.103514},
    eprint = {0912.1943},
    archivePrefix = {arXiv},
    primaryClass = {astro-ph.CO}
}

@article{Jullo2007,
    author = {Jullo, E. and Kneib, J.-P. and Limousin, M. and {El\'iasd\'ottir}, \'A. and Marshall, P. J. and Verdugo, T.},
    title = {A Bayesian approach to strong lensing modelling of galaxy clusters},
    journal = {New Journal of Physics},
    year = {2007},
    volume = {9},
    pages = {447},
    doi = {10.1088/1367-2630/9/12/447},
    note = {Publicado en el Número Especial "Gravitational Lensing". El código \textsc{Lenstool} está disponible en: \url{http://www.oamp.fr/cosmology/lenstool/}}
}

@article{akaike1974,
  title={A new look at the statistical model identification},
  author={Akaike, Hirotugu},
  journal={IEEE transactions on automatic control},
  volume={19},
  number={6},
  pages={716--723},
  year={1974},
  publisher={IEEE}
}

@article{schwarz1978,
  title={Estimating the dimension of a model},
  author={Schwarz, Gideon},
  journal={The annals of statistics},
  volume={6},
  number={2},
  pages={461--464},
  year={1978},
  publisher={Institute of Mathematical Statistics}
}

@ARTICLE{2019MNRAS.482.1858Y,
       author = {{Yang}, Weiqiang and {Pan}, Supriya and {Xu}, Lixin and {Mota}, David F.},
        title = "{Effects of anisotropic stress in interacting dark matter - dark energy scenarios}",
      journal = {Monthly Notices of the Royal Astronomical Society},
         year = 2019,
        month = jan,
       volume = {482},
       number = {2},
        pages = {1858-1871},
          doi = {10.1093/mnras/sty2789},
archivePrefix = {arXiv},
       eprint = {1804.08455},
 primaryClass = {astro-ph.CO},
       adsurl = {https://ui.adsabs.harvard.edu/abs/2019MNRAS.482.1858Y}
}

@ARTICLE{2025PDU....4801951G,
       author = {{Goswami}, Sangita and {Das}, Sudipta},
        title = "{Exploring parametrized dark energy models in interacting scenario}",
      journal = {Physics of the Dark Universe},
         year = 2025,
        month = may,
       volume = {48},
          eid = {101951},
        pages = {101951},
          doi = {10.1016/j.dark.2025.101951},
archivePrefix = {arXiv},
       eprint = {2505.16438},
 primaryClass = {gr-qc},
       adsurl = {https://ui.adsabs.harvard.edu/abs/2025PDU....4801951G}
}

@ARTICLE{2008RPPh...71e6901L,
       author = {{Linder}, Eric V.},
        title = "{Mapping the cosmological expansion}",
      journal = {Reports on Progress in Physics},
         year = 2008,
        month = may,
       volume = {71},
       number = {5},
          eid = {056901},
        pages = {056901},
          doi = {10.1088/0034-4885/71/5/056901},
archivePrefix = {arXiv},
       eprint = {0801.2968},
 primaryClass = {astro-ph},
       adsurl = {https://ui.adsabs.harvard.edu/abs/2008RPPh...71e6901L}
}

@ARTICLE{2012JCAP...03..016C,
       author = {{Cao}, Shuo and {Pan}, Yu and {Biesiada}, Marek and {Godlowski}, Wlodzimierz and {Zhu}, Zong-Hong},
        title = "{Constraints on cosmological models from strong gravitational lensing systems}",
      journal = {Journal of Cosmology and Astroparticle Physics},
         year = 2012,
        month = mar,
       volume = {2012},
       number = {3},
          eid = {016},
        pages = {016},
          doi = {10.1088/1475-7516/2012/03/016},
archivePrefix = {arXiv},
       eprint = {1105.6226},
 primaryClass = {astro-ph.CO},
       adsurl = {https://ui.adsabs.harvard.edu/abs/2012JCAP...03..016C}
}

@ARTICLE{2015ApJ...813...69M,
       author = {{Maga{\~n}a}, Juan and {Motta}, V. and {C{\'a}rdenas}, V{\'\i}ctor H. and {Verdugo}, T. and {Jullo}, Eric},
        title = "{A Magnified Glance into the Dark Sector: Probing Cosmological Models with Strong Lensing in A1689}",
      journal = {The Astrophysical Journal},
         year = 2015,
        month = nov,
       volume = {813},
       number = {1},
          eid = {69},
        pages = {69},
          doi = {10.1088/0004-637X/813/1/69},
archivePrefix = {arXiv},
       eprint = {1509.08162},
 primaryClass = {astro-ph.CO},
       adsurl = {https://ui.adsabs.harvard.edu/abs/2015ApJ...813...69M}
}

@ARTICLE{2004PhRvD..70d3534L,
       author = {{Linder}, Eric V.},
        title = "{Strong gravitational lensing and dark energy complementarity}",
      journal = {Physical Review D},
         year = 2004,
        month = aug,
       volume = {70},
       number = {4},
          eid = {043534},
        pages = {043534},
          doi = {10.1103/PhysRevD.70.043534},
archivePrefix = {arXiv},
       eprint = {astro-ph/0401433},
 primaryClass = {astro-ph},
       adsurl = {https://ui.adsabs.harvard.edu/abs/2004PhRvD..70d3534L}
}

@ARTICLE{2011PhRvD..84l3529L,
       author = {{Linder}, Eric V.},
        title = "{Lensing time delays and cosmological complementarity}",
      journal = {Physical Review D},
         year = 2011,
        month = dec,
       volume = {84},
       number = {12},
          eid = {123529},
        pages = {123529},
          doi = {10.1103/PhysRevD.84.123529},
archivePrefix = {arXiv},
       eprint = {1109.2592},
 primaryClass = {astro-ph.CO},
       adsurl = {https://ui.adsabs.harvard.edu/abs/2011PhRvD..84l3529L}
}

@ARTICLE{2026JHEAp..5100534Z,
       author = {{Zhu}, Ziyan and {Jiang}, Qingquan and {Liu}, Yu and {Wu}, Puxun and {Liang}, Nan},
        title = "{Cosmological constraints on the phenomenological interacting dark energy model with Fermi gamma-ray bursts and DESI DR2}",
      journal = {Journal of High Energy Astrophysics},
         year = 2026,
        month = mar,
       volume = {51},
          eid = {100534},
        pages = {100534},
          doi = {10.1016/j.jheap.2025.100534},
archivePrefix = {arXiv},
       eprint = {2511.16032},
 primaryClass = {astro-ph.CO},
       adsurl = {https://ui.adsabs.harvard.edu/abs/2026JHEAp..5100534Z}
}

@ARTICLE{2026NuPhB102517368S,
       author = {{Saha}, Somnath and {Saha}, Subhajit and {Mahata}, Nilanjana},
        title = "{The peculiar case of the Viaggiu holographic dark energy}",
      journal = {Nuclear Physics B},
         year = 2026,
        month = apr,
       volume = {1025},
          eid = {117368},
        pages = {117368},
          doi = {10.1016/j.nuclphysb.2026.117368},
archivePrefix = {arXiv},
       eprint = {2603.15067},
 primaryClass = {gr-qc},
       adsurl = {https://ui.adsabs.harvard.edu/abs/2026NuPhB102517368S}
}

@ARTICLE{2026arXiv260315178H,
       author = {{Halder}, Amlan K. and {Paliathanasis}, Andronikos and {Viaggiu}, Stefano and {Al Mamon}, Abdulla and {Saha}, Subhajit},
        title = "{Viaggiu holographic dark energy in light of DESI DR2}",
      journal = {arXiv e-prints},
         year = 2026,
        month = mar,
          eid = {arXiv:2603.15178},
        pages = {arXiv:2603.15178},
          doi = {10.48550/arXiv.2603.15178},
archivePrefix = {arXiv},
       eprint = {2603.15178},
 primaryClass = {gr-qc},
       adsurl = {https://ui.adsabs.harvard.edu/abs/2026arXiv260315178H}
}

@ARTICLE{2026arXiv260423992V,
       author = {{Verdugo}, Tom{\'a}s and {Hern{\'a}ndez-Almada}, Alberto and {Garc{\'\i}a-Aspeitia}, Miguel A. and {Maga{\~n}a}, Juan and {Motta}, Ver{\'o}nica},
        title = "{Reconstructing the cosmic expansion with a generalized q(z) parameterization: A decelerating Universe from late-time constraints}",
      journal = {arXiv e-prints},
         year = 2026,
        month = apr,
          eid = {arXiv:2604.23992},
        pages = {arXiv:2604.23992},
archivePrefix = {arXiv},
       eprint = {2604.23992},
 primaryClass = {astro-ph.CO},
       adsurl = {https://ui.adsabs.harvard.edu/abs/2026arXiv260423992V}
}

@ARTICLE{2024RPPh...87c6901W,
       author = {{Wang}, B. and {Abdalla}, E. and {Atrio-Barandela}, F. and {Pav{\'o}n}, D.},
        title = "{Further understanding the interaction between dark energy and dark matter: current status and future directions}",
      journal = {Reports on Progress in Physics},
         year = 2024,
        month = mar,
       volume = {87},
       number = {3},
          eid = {036901},
        pages = {036901},
          doi = {10.1088/1361-6633/ad2527},
archivePrefix = {arXiv},
       eprint = {2402.00819},
 primaryClass = {astro-ph.CO},
       adsurl = {https://ui.adsabs.harvard.edu/abs/2024RPPh...87c6901W}
}

@ARTICLE{2017JCAP...01..028C,
       author = {{Costa}, Andr{\'e} A. and {Xu}, Xiao-Dong and {Wang}, Bin and {Abdalla}, E.},
        title = "{Constraints on interacting dark energy models from Planck 2015 and redshift-space distortion data}",
      journal = {Journal of Cosmology and Astroparticle Physics},
         year = 2017,
        month = jan,
       volume = {2017},
       number = {1},
          eid = {028},
        pages = {028},
          doi = {10.1088/1475-7516/2017/01/028},
archivePrefix = {arXiv},
       eprint = {1605.04138},
 primaryClass = {astro-ph.CO},
       adsurl = {https://ui.adsabs.harvard.edu/abs/2017JCAP...01..028C}
}

@ARTICLE{2026PhRvD.113b3515P,
       author = {{Pan}, Supriya and {Paul}, Sivasish and {Saridakis}, Emmanuel N. and {Yang}, Weiqiang},
        title = "{Interacting dark energy after DESI DR2: A challenge for the {\ensuremath{\Lambda}}CDM paradigm?}",
      journal = {Physical Review D},
         year = 2026,
        month = jan,
       volume = {113},
       number = {2},
          eid = {023515},
        pages = {023515},
          doi = {10.1103/5y21-k39n},
archivePrefix = {arXiv},
       eprint = {2504.00994},
 primaryClass = {astro-ph.CO},
       adsurl = {https://ui.adsabs.harvard.edu/abs/2026PhRvD.113b3515P}
}

@ARTICLE{2023JCAP...10..047D,
       author = {{Das}, Santanu and {Nasiri}, Arad and {Yazdi}, Yasaman K.},
        title = "{Aspects of Everpresent {\ensuremath{\Lambda}}. Part I. A fluctuating cosmological constant from spacetime discreteness}",
      journal = {Journal of Cosmology and Astroparticle Physics},
         year = 2023,
        month = oct,
       volume = {2023},
       number = {10},
          eid = {047},
        pages = {047},
          doi = {10.1088/1475-7516/2023/10/047},
archivePrefix = {arXiv},
       eprint = {2304.03819},
 primaryClass = {gr-qc},
       adsurl = {https://ui.adsabs.harvard.edu/abs/2023JCAP...10..047D}
}

@ARTICLE{2025arXiv250524732C,
       author = {{Cai}, Yifu and {Ren}, Xin and {Qiu}, Taotao and {Li}, Mingzhe and {Zhang}, Xinmin},
        title = "{The Quintom theory of dark energy after DESI DR2}",
      journal = {arXiv e-prints},
         year = 2025,
        month = may,
          eid = {arXiv:2505.24732},
        pages = {arXiv:2505.24732},
          doi = {10.48550/arXiv.2505.24732},
archivePrefix = {arXiv},
       eprint = {2505.24732},
 primaryClass = {astro-ph.CO},
       adsurl = {https://ui.adsabs.harvard.edu/abs/2025arXiv250524732C}
}

@ARTICLE{2022JCAP...04..004L,
       author = {{Lee}, Bum-Hoon and {Lee}, Wonwoo and {{\'O} Colg{\'a}in}, Eoin and {Sheikh-Jabbari}, M.~M. and {Thakur}, Somyadip},
        title = "{Is local H $_{0}$ at odds with dark energy EFT?}",
      journal = {Journal of Cosmology and Astroparticle Physics},
         year = 2022,
        month = apr,
       volume = {2022},
       number = {4},
          eid = {004},
        pages = {004},
          doi = {10.1088/1475-7516/2022/04/004},
archivePrefix = {arXiv},
       eprint = {2202.03906},
 primaryClass = {astro-ph.CO},
       adsurl = {https://ui.adsabs.harvard.edu/abs/2022JCAP...04..004L}
}

@ARTICLE{2026PDU....5202268O,
       author = {{{\'O} Colg{\'a}in}, Eoin and {Pourojaghi}, Saeed and {Sheikh-Jabbari}, M.~M. and {Yin}, Lu},
        title = "{How much has DESI dark energy evolved since DR1?}",
      journal = {Physics of the Dark Universe},
         year = 2026,
        month = jun,
       volume = {52},
          eid = {102268},
        pages = {102268},
          doi = {10.1016/j.dark.2026.102268},
archivePrefix = {arXiv},
       eprint = {2504.04417},
 primaryClass = {astro-ph.CO},
       adsurl = {https://ui.adsabs.harvard.edu/abs/2026PDU....5202268O}
}

@article{PhysRevD.110.123519,
  title = {Nonparametric late-time expansion history reconstruction and implications for the Hubble tension in light of recent DESI and type Ia supernovae data},
  author = {Jiang, Jun-Qian and Pedrotti, Davide and da Costa, Simony Santos and Vagnozzi, Sunny},
  journal = {Phys. Rev. D},
  volume = {110},
  issue = {12},
  pages = {123519},
  numpages = {18},
  year = {2024},
  month = {Dec},
  publisher = {American Physical Society},
  doi = {10.1103/PhysRevD.110.123519},
  url = {https://link.aps.org/doi/10.1103/PhysRevD.110.123519}
}



\end{document}